\documentclass{aastex62}
\usepackage{lineno}
\begin{document}

\title{The spatial distributions of blue main-sequence stars in Magellanic Cloud star clusters} 
\author[0000-0002-3180-2327]{Yujiao Yang}
\affiliation{Kavli Institute for Astronomy \& Astrophysics, Peking
  University, Yi He Yuan Lu 5, Hai Dian District, Beijing 100871,
  China} 
\affiliation{Department of Astronomy, School of Physics, Peking
  University, Yi He Yuan Lu 5, Hai Dian District, Beijing 100871,
  China}
\affiliation{Department of Physics and Astronomy, Macquarie
  University, Balaclava Road, Sydney, NSW 2109, Australia}
\affiliation{Research Centre for Astronomy, Astrophysics and
  Astrophotonics, Macquarie University, Balaclava Road, Sydney, NSW
  2109, Australia}

\author[0000-0002-3084-5157]{Chengyuan Li} 
\affiliation{School of Physics and Astronomy, Sun Yat-sen University,
  Zhuhai 519082, China} 
\affiliation{Department of Physics and Astronomy, Macquarie
  University, Balaclava Road, Sydney, NSW 2109, Australia}
\affiliation{Research Centre for Astronomy, Astrophysics and
  Astrophotonics, Macquarie University, Balaclava Road, Sydney, NSW
  2109, Australia}
\affiliation{Key Laboratory for Optical Astronomy, National
  Astronomical Observatories, Chinese Academy of Sciences, 20A Datun
  Road, Chaoyang District, Beijing 100012, China}

\author[0000-0002-7203-5996]{Richard de Grijs}
\affiliation{Department of Physics and Astronomy, Macquarie
  University, Balaclava Road, Sydney, NSW 2109, Australia}
\affiliation{Research Centre for Astronomy, Astrophysics and
  Astrophotonics, Macquarie University, Balaclava Road, Sydney, NSW
  2109, Australia}

\author[0000-0001-9073-9914]{Licai Deng}
\affiliation{Key Laboratory for Optical Astronomy, National
  Astronomical Observatories, Chinese Academy of Sciences, 20A Datun
  Road, Chaoyang District, Beijing 100012, China}
\affiliation{School of Astronomy and Space Science, University of the
  Chinese Academy of Sciences, Huairou 101408, China}
\affiliation{Department of Astronomy, China West Normal University,
  Nanchong 637002, China}
  
\correspondingauthor{Yujiao Yang; Chengyuan Li}
\email{yujiaoyang@pku.edu.cn; lichengy5@mail.sysu.edu.cn}

\begin{abstract}
The color--magnitude diagrams (CMDs) of young star clusters show that,
particularly at ultraviolet wavelengths, their upper main sequences
(MSs) bifurcate into a sequence comprising the bulk population and a
blue periphery. The spatial distribution of stars is crucial to
understand the reasons for these distinct stellar populations. This
study uses high-resolution photometric data obtained with the {\sl
  Hubble Space Telescope} to study the spatial distributions of the
stellar populations in seven Magellanic Cloud star clusters. The
cumulative radial number fractions of blue stars within four clusters
are strongly anti-correlated with those of the high-mass-ratio
binaries in the bifurcated region, with negative Pearson coefficients
$< -0.7$. Those clusters generally are young or in an early dynamical
evolutionary stage. In addition, a supporting $N$-body simulation
suggests the increasing percentage of blue-MS stars from the cluster
centers to their outskirts may be associated with the dissolution of
soft binaries. This study provides a different perspective to explore
the MS bimodalities in young clusters and adds extra puzzles. A more
comprehensive study combined with detailed simulations is needed in
the future.
\end{abstract}


\section{Introduction} \label{sec : intro}

In recent years, unexpected features in star cluster color--magnitude
diagrams (CMDs) have challenged the traditional concept that star
clusters were formed as simple stellar populations, i.e., with member
stars that share the same age and the same metallicity (within some
tolerance). For example, evidence from both photometric and
spectrometric observations reveals that extended main-sequence
turnoffs (eMSTOs) are common in almost all intermediate-age
($\sim1$--2 Gyr old) Magellanic Cloud (MC) clusters
\citep[e.g.,][]{2007MNRAS.379..151M, Milone2009, Girardi2013, Li2016}
and also in some young massive MC clusters
\citep[e.g.,][]{Milone2015multiple, Milone2016multiple,
  Milone2017mnras, Li2017}. The reason behind those observed features
is still subject to debate.

Clusters younger than 400 Myr usually feature split main sequences
(MSs), especially when the observations include ultraviolet (UV) bands
\citep{Milone2018}. In this case, the bulk of the cluster stars
populate the redder sequence (with a red tail caused by binary
systems) and a smaller fraction of stars populate the blue
periphery. \cite{2013A&A...555A.143M} first observed a split MS in NGC
1844, a $\sim$150 Myr-old Large Magellanic Cloud (LMC)
cluster. Subsequent studies \citep[e.g.,][]{2015MNRAS.453.2637D,
  Milone2015multiple, Milone2016multiple, Milone2017mnras} confirmed
that bifurcated MSs are an intrinsic feature of star cluster
CMDs. They cannot be interpreted as abundance anomalies, differential
reddening, photometric uncertainties, or contamination by field stars.

Based on comparisons of observed CMDs and stellar evolution
  models, multiple studies show that different stellar rotation rates
  can mimic the observed eMSTOs \citep[e.g.,][]{2009MNRAS.398L..11B,
    2019A&A...622A..66G} and split MSs
  \citep[e.g.,][]{2015MNRAS.453.2637D,Milone2016multiple}. Two main
  stellar rotation effects determine stellar CMD loci: (i) the
  centrifugal force induced by stellar rotation causes temperature
  differences across the stellar surface, thus resulting in different
  colors under different viewing angles; (ii) rapid rotation enhances
  interior mixing of stellar material, which changes the stellar
  evolutionary path in the CMD. Observed blue- and red-MS stars can
be largely reproduced by assuming a coeval population composed of
$\sim$30\% slowly or non-rotating stars and $\sim$70\% rapidly
rotating stars \citep[defined as $\Omega_{\rm ini}/ \Omega_{\rm crit}
  > 0.9$, where $\Omega_{\rm crit}$ is the break-up angular velocity;
  e.g.,][]{Milone2016multiple, Milone2017mnras,
  2017MNRAS.467.3628C}. Color variations of stars characterized by
different rotation rates observed in spectroscopic studies
\citep[][]{2017ApJ...846L...1D, 2018MNRAS.480.1689K,
  2018MNRAS.480.3739B, 2018ApJ...863L..33M, 2018AJ....156..116M,
  2019ApJ...876..113S, 2020MNRAS.492.2177K} and the presence of a
large number of Be stars in at least some clusters
\citep[]{2017MNRAS.465.4795B, Milone2018} support the notion that a
spread in stellar rotation rates could be the main cause of the
observed split MSs and eMSTOs in young clusters.

Nevertheless, the reason behind the bimodality of rotation rates is
not yet clear. We know that most massive stars are characterized by
high rotation rates \citep{2013A&A...550A.109D}, and their masses are
consistent with the masses of middle and upper-MS stars in young star
clusters. Then the question arises as to why young clusters host large
numbers of slowly or non-rotating massive
stars. \cite{2015MNRAS.453.2637D} suggested that all stars are
initially rapid rotators. Binary interactions would slow down some of
them, resulting in the observed bimodal distribution of rotation
rates. \cite{2019ApJ...876..113S} also suggested that tidal locking in
binary systems might induce a stellar rotation dichotomy. However,
\cite{2020MNRAS.495.1978B} recently suggested that the bimodal
rotation distribution may be established during the early cluster
formation stage, and the bimodality can be maintained throughout the
pre-MS and MS lifetime. Studying the spatial distributions of blue-MS
stars may help us narrow down the origin of the dichotomy.

Several studies have analyzed the radial distributions of blue-MS
stars. For instance, the fraction of blue-MS stars in NGC 1866
increases significantly from the cluster center to its periphery
\citep[][their Fig. 9]{Milone2017mnras}, indicating that blue-MS stars
are less concentrated than the MS stars in general. In NGC 1856, the
population ratio of blue- to red-MS stars remains approximately
unchanged within the cluster field, indicating a homogeneous spatial
distribution \citep{Li2017}. \cite{Yang2018} found that the cumulative
fraction of blue-MS stars among the full sample of MS stars in NGC
1850 increases from the cluster center to its outer regions, which is
strongly anti-correlated with the radial profile of the cluster's
high-mass-ratio binaries. Here we present a follow-up study focusing
on the blue-MS stars' spatial distributions in a larger cluster
sample.

In this paper, we selected seven MC star clusters featuring split MSs
(NGC 1755, NGC 1805, NGC 1818, NGC 1850, NGC1866, NGC 2164, and NGC
330) to study the spatial distributions of their stellar
components. These clusters have ages between $\sim$40 Myr and
$\sim$250 Myr and high masses ($\log (M/M_{\odot})
\sim3.5$--5.0). \cite{Milone2018} showed that all of these clusters
have distinct dual MSs. An additional $N$-body simulation was run to
study changes in the binaries' spatial distribution in an isolated
star cluster.

This paper is organized as follows. Section~\ref{sec : data} presents
the observational data, the fitting processes employed to obtain the
cluster parameters, and the methodology to select distinct stellar
populations. The results are described in Section~\ref{sec :
  obsresult}, followed by a discussion in Section~\ref{sec : dis} and
a summary in Section~\ref{sec : sum}.

\section{Data and data analysis} \label{sec : data}

The raw images analyzed here were observed with the Ultraviolet and
Visual Channel of the Wide Field Camera 3 (UVIS/WFC3) on board the
{\sl Hubble Space Telescope} ({\sl HST}). The observational details
are summarized in Table~\ref{table : data}. Point-spread-function
photometry was performed using the WFC3 modules in the DOLPHOT
package\footnote{DOLPHOT is a stellar photometry package for analysis
  of {\sl HST} data; see \url{http://
    americano.dolphinism.com/dolphot.}}.

Flat-field images (with suffix `flt.fits') of NGC 1755 and NGC 1866
were collected from the same program; they had already been drizzled
by the WFC3 reduction pipeline. High-quality photometric data can be
obtained by following the process steps recommended in the {\it
  DOLPHOT/WFC3 User's Guide}. The main steps include running the
\textit{wfc3mask}, \textit{splitgroups}, \textit{calcsky}, and
\textit{dolphot} tasks.

For the other five clusters, raw images in the F336W and F814W filters
were collected from different programs, so we have to be mindful that
the variations in the pointings may cause spatial
offsets. DrizzlePac\footnote{DrizzlePac is a software package designed
  to align and combine {\sl HST} images. The task
  \textit{astrodrizzle} is part of the pipeline used to generate
  Barbara A. Mikulski Archive for Space Telescopes (MAST) data; see
  \url{http://drizzlepac.stsci.edu}.} can align and combine
images. Post-pipeline image processing was applied to those clusters
so as to obtain high-quality photometric data.

Specifically, we first used the task \textit{tweakreg} to align the
images obtained for a given filter and update the World Coordinate
System (WCS) information stored in the header of each image. We used
the task \textit{astrodrizzle} to combine the images and produce a new
drizzled product (with suffix `drz.fits') for each filter. Next, we
improved the alignment between the drizzled F336W and F814W images by
employing \textit{tweakreg}. The updated WCS information stored in
both `drz.fits' images minimizes the offset between the observation
programs, which was propagated back to the original `flt.fits' images
by the task \textit{tweakback}. For all DOLPHOT output data, we
applied the same criteria to select high-quality photometric data,
i.e., $\textit{Signal-to-noise} > 10$, $|\textit{Object sharpness} | <
0.2$, $\textit{Crowding} < 0.5$, $\textit{Object type} = 1$ (good
stars), and $\textit{Photometry quality flag} = 0$. The
  sharpness cuts were chosen so as to remove contamination by cosmic
  rays (positive sharpness) and clusters or galaxies (negative
  sharpness). The crowding parameter is the difference between the
  magnitude of a star measured including or excluding contamination by
  nearby stars in the image; it is zero for an isolated star. Our
  crowding cut was aimed at removing stars whose photometry was
  significantly affected by a high stellar density. The values adopted
  for \textit{Object type} and \textit{Photometry quality flag}
  correspond to good stars and perfectly recovered stars,
  respectively. Stars satisfying those criteria were selected as the
final stellar catalog for each cluster.

\begin{table}[tb]
\caption{Details about the raw UVIS/WFC3 images used.}
\centering
\begin{tabular}{c|ccc|cc}
\hline
\hline
Cluster & Proposal ID & PI Name & Filter & Root Name & Exposure Time (s) \\ [1ex]
\hline
{NGC 1755}& {14204} & {A.P. Milone  } & {F814W} & icu807ctq & 90  \\
{       } & {     } & {             } & {     } & icu807d3q & 678 \\
{       } & {14204} & {             } & {F336W} & icu807cvq & 711 \\
{       } & {     } & {             } & {     } & icu807czq & 711 \\
\hline
{NGC 1805}& {14710} & {A.P. Milone  } & {F814W} & id6i02ftq & 90  \\
{       } & {     } & {             } & {     } & id6i02g3q & 666 \\
{       } & {13727} & {J. Kalirai   } & {F336W} & ick002omq & 100 \\
{       } & {     } & {             } & {     } & ick002p8q & 947 \\
\hline
{NGC 1818}& {14710} & {A.P. Milone  } & {F814W} & id6i01c3q & 90  \\
{       } & {     } & {             } & {     } & id6i01cdq & 666 \\
{       } & {13727} & {J. Kalirai   } & {F336W} & ick004wpq & 100 \\
{       } & {     } & {             } & {     } & ick004wkq & 790 \\
\hline
{NGC 1850}& {14714} & {P. Goudfrooij} & {F814W} & icza01b5q & 350 \\
{       } & {     } & {             } & {     } & icza01bfq & 440 \\
{       } & {14069} & {N. Bastian   } & {F336W} & icz601bcq & 370 \\
{       } & {     } & {             } & {     } & icz601bqq & 260 \\
\hline
{NGC 1866}& {14204} & {A.P. Milone  } & {F814W} & icu804imq & 90  \\
{       } & {     } & {             } & {     } & icu804juq & 678 \\
{       } & {14204} & {             } & {F336W} & icu804irq & 711 \\
{       } & {     } & {             } & {     } & icu804jqq & 711 \\
\hline
{NGC 2164}& {14710} & {A.P. Milone  } & {F814W} & id6i04eoq & 90  \\
{       } & {     } & {             } & {     } & id6i04eyq & 758 \\
{       } & {13727} & {J. Kalirai   } & {F336W} & ick001h0q & 100 \\
{       } & {     } & {             } & {     } & ick001gwq & 790 \\
\hline
{NGC 330} & {14710} & {A.P. Milone  } & {F814W} & id6i03aeq & 90  \\
{       } & {     } & {             } & {     } & id6i03apq & 680 \\
{       } & {13727} & {J. Kalirai   } & {F336W} & ick003mlq & 100 \\
{       } & {     } & {             } & {     } & ick003mhq & 805 \\ [1ex]
\hline
\end{tabular}
\label{table : data}
\end{table}

Basic parameters of the clusters were determined as follows. First, to
determine the coordinates of the cluster center, we calculated a
number-density contour map in the R.A.--Dec plane and adopted the
location where the number densities approach the largest value as the
center. Then we divided stars into annular rings and calculated the
corresponding surface brightness in F814W. Core and cluster radii were
derived using a surface brightness fitting process. Since almost all
rich LMC clusters studied by \cite{Elson1987} extend beyond their
tidal radii, we applied a more suitable `Elson, Fall, \& Freeman'
(EFF) model \citep{Elson1987} to fit the surface brightness
profile. The EFF model obeys
\begin{equation}
\mu(r) = \mu_{0}(1+\frac{r^2}{a^2})^{- \gamma /2},
\end{equation}
where $\mu_0$ is the central surface brightness, $a$ is a measure of
the core radius, and $\gamma$ is the power-law index. The core radius
$r_{\rm c}$ pertaining to the standard King model can be obtained
easily,
\begin{equation}
r_{\rm c} = a(2^{2/\gamma} -1)^{1/2}.
\end{equation}

Matching theoretical and observational profiles becomes unreliable in
the radially outermost annular rings because of large
uncertainties. Therefore, we adopted the position where the cluster's
surface brightness profile reaches the average field level as the
cluster area, which was determined by eye. We specifically focus on
the ratio of populations, so that small variations in the overall
cluster area have a negligible impact on our final results. Cluster
ages, distance moduli, and reddening values were derived from
isochrone fits. The minimum-$\chi^2$ method
\citep{2016ApJ...823...18C} was applied to obtain the best-fitting
isochrones from the PARSEC model suite \citep{Bressan2012}. For
details, see Section~\ref{subsection : isochrone fitting}. The
resulting parameters are included in Table~\ref{table : parameter}.
We realize that using non-rotating isochrones to fit the dominant
population (the red-MS stars) could potentially increase the
uncertainties in the resulting parameters, since the red-MS stars are
generally thought to be rapidly rotating stars. These parameters will
be used to estimate the clusters' dynamical ages (defined as the ratio
of a cluster's chronological and half-mass relaxation timescales,
${t_{\rm iso}}/{t_{\rm rh}}$). The resulting order of the dynamical
ages determined in this paper is the same as that estimated by
\cite{2005ApJS..161..304M} (except for NGC 1755, which those authors
did not consider). Therefore, we conclude that our approach is robust
and that the derived isochrone parameters have a negligible impact on
our final results.
  
\begin{table}[!htb]
\caption{Parameters of the seven MC clusters derived here. (1) Cluster
  name, (2) Host galaxy, (3) Right Ascension (J2000), (4) Declination
  (J2000), (5) Age, (6) Distance modulus, (7) Reddening, (8)
  Best-fitting core radius, (9) Half-mass radius, (10) Total cluster
  mass. Core radii were determined from the best-fitting EFF model in
  the F814W band.}  \centering
\begin{tabular}{cc|cc|ccc|ccc}
\hline
\hline
Cluster & Galaxy & $\alpha_{\rm J2000}$ & $\delta_{\rm J2000}$ & Age & $(m-M)_{\rm 0}$ & $A_{\rm V}$ & $R_{\rm c}$ & $R_{\rm h}$ & $\log (M/M_{\odot})$ \\
  & & & & (Myr) & (mag) & (mag) & (arcsec) & (arcsec) & \\       
(1) & (2) & (3) & (4) & (5) & (6) & (7) & (8) & (9) & (10) \\[1ex]
\hline
NGC 1755 & LMC & $04^{\rm h} 55^{\rm m} 15.16^{\rm s}$ & $-68^{\circ} 12' 18.44''$ & 127.0$\pm87.12$ & 18.20$\pm0.29$ & 0.34$\pm0.11$ & 7.83 $\pm1.35$ & 10.32 & $4.161^{+0.097}_{-0.056}$ \\
NGC 1805 & LMC & $05^{\rm h} 02^{\rm m} 21.66^{\rm s}$ & $-66^{\circ} 06' 41.76''$ & 44.0  $\pm40.09$ & 18.20$\pm0.32$ & 0.29$\pm0.20$ & 3.77 $\pm0.53$ & 5.15   & $3.949^{+0.619}_{-0.018}$ \\
NGC 1818 & LMC & $05^{\rm h} 04^{\rm m} 13.85^{\rm s}$ & $-66^{\circ} 26' 02.16''  $ & 48.0  $\pm20.86$ & 18.29$\pm0.38$ & 0.30$\pm0.15$ & 8.50 $\pm1.09$ & 9.12   & $4.198^{+0.13}_{-0.071}$ \\
NGC 1850 & LMC & $05^{\rm h} 08^{\rm m} 45.17^{\rm s}$ & $-68^{\circ} 45' 42.78''$ & 86.0  $\pm45.59$ & 18.30$\pm0.45$ & 0.50$\pm0.13$ &13.21$\pm3.79$ & 13.63 & $4.717^{+0.165}_{-0.152}$ \\
NGC 1866 & LMC & $05^{\rm h} 13^{\rm m} 38.70^{\rm s}$ & $-65^{\circ} 27' 52.15''$ & 226.0$\pm54.79$ & 18.20$\pm0.37$ & 0.45$\pm0.12$ &15.37$\pm5.39$ & 13.92 & $4.484^{+0.114}_{-0.062}$ \\
NGC 2164 & LMC & $05^{\rm h} 58^{\rm m} 55.92^{\rm s}$ & $-68^{\circ} 30' 57.38''$ & 153.0$\pm55.96$ & 18.20$\pm0.36$ & 0.24$\pm0.12$ & 7.04 $\pm0.52$ & 7.15   & $4.06^{+0.152}_{-0.111}$ \\
NGC 330  & SMC & $00^{\rm h} 56^{\rm m} 18.56^{\rm s}$ & $-72^{\circ} 27' 48.74''$ & 44.0 $\pm39.39$ & 18.60$\pm0.39$  & 0.37$\pm0.13$ & 9.10 $\pm2.93$ & 13.19 & $4.385^{+0.134}_{-0.813}$ \\[1ex]
\hline
\end{tabular}
 {\raggedright Note: Cluster ages, distance moduli, and reddening
   values were derived from the best-fitting non-rotating
   isochrone. \par}
\label{table : parameter}
\end{table}

\begin{figure}[!htb]
\centering
\includegraphics[width = 0.8 \textwidth]{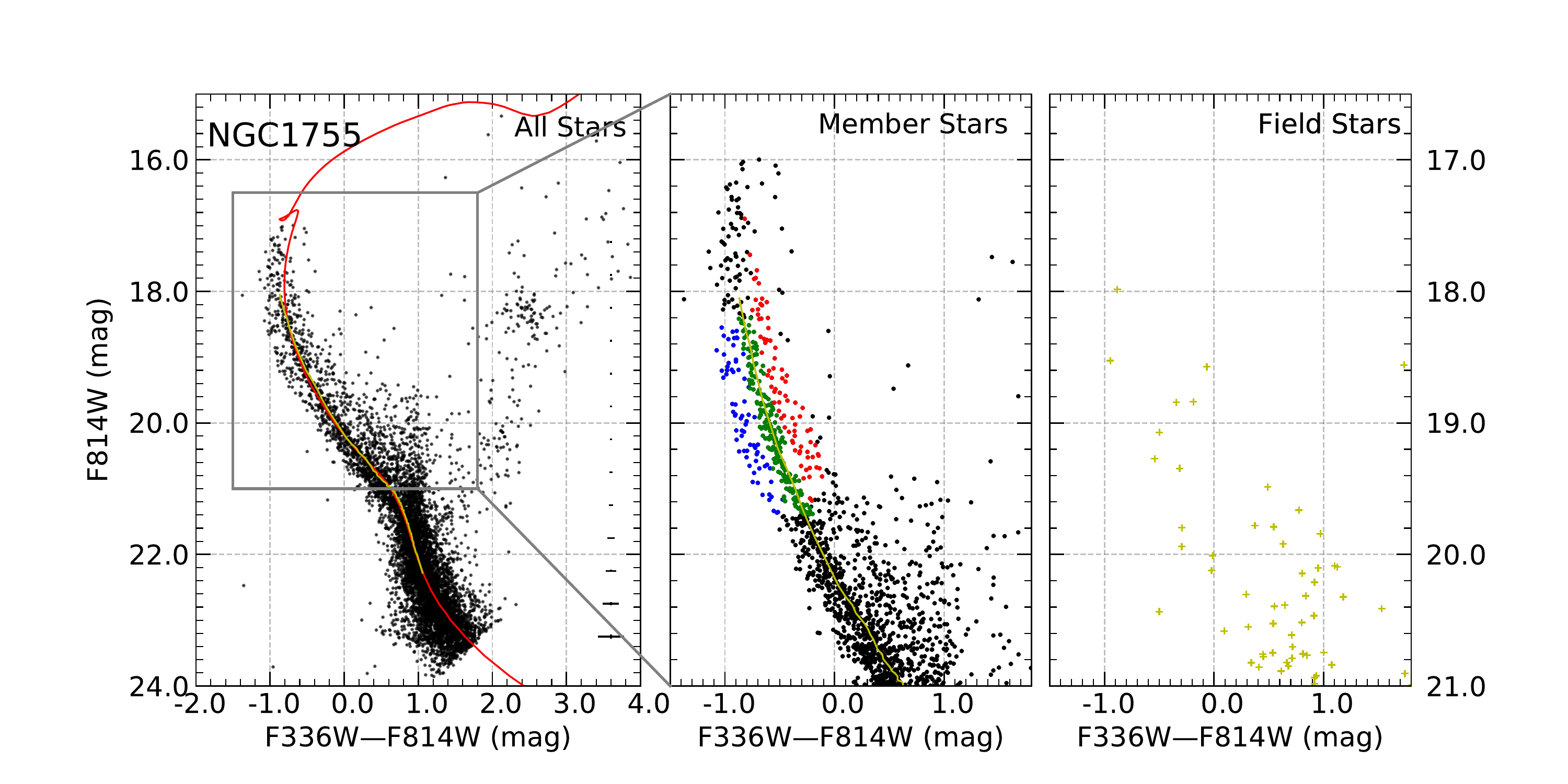}
\vspace{0cm}
\includegraphics[width = 0.8 \textwidth]{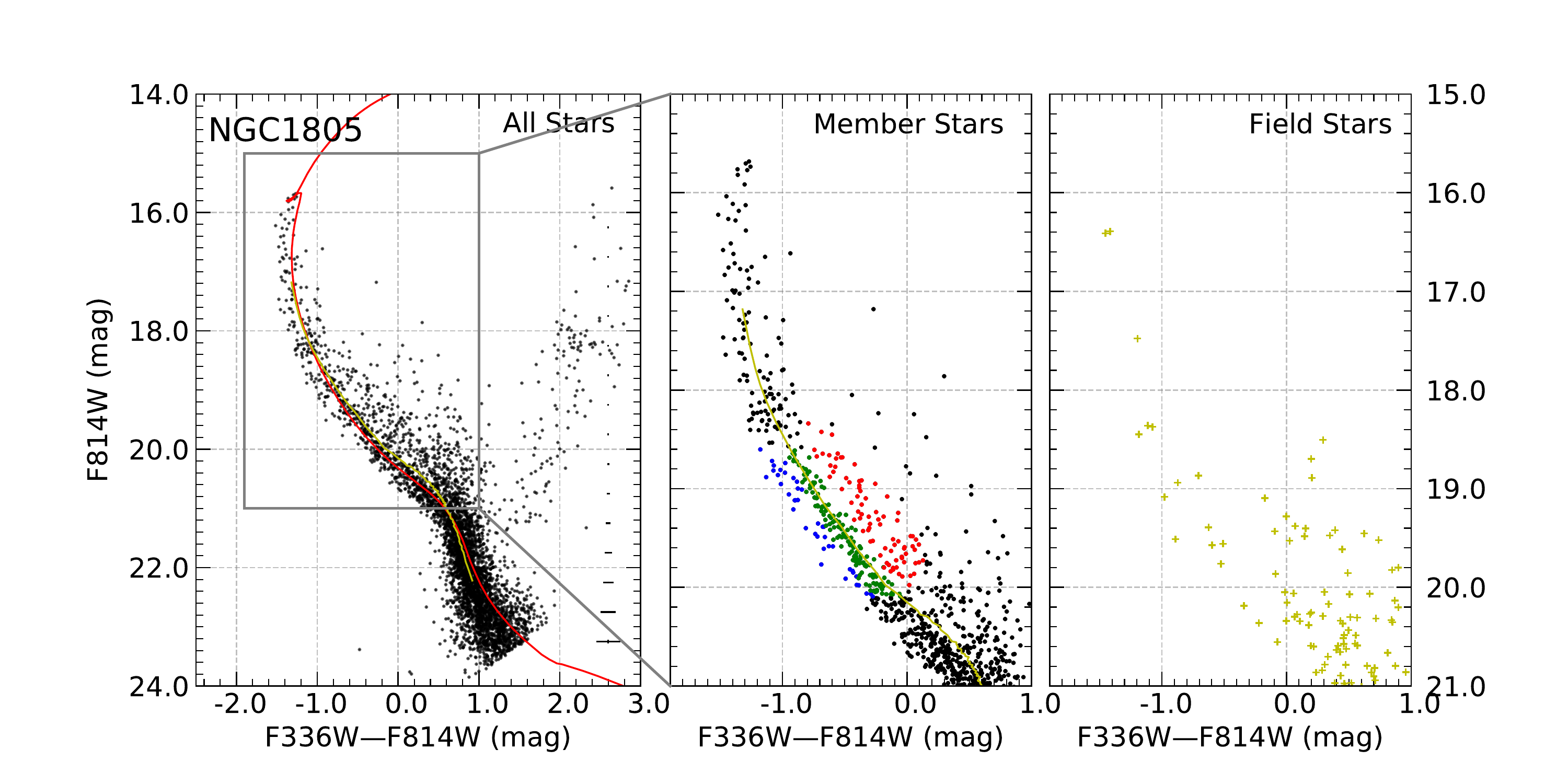}
\vspace{0cm}
\includegraphics[width = 0.8 \textwidth]{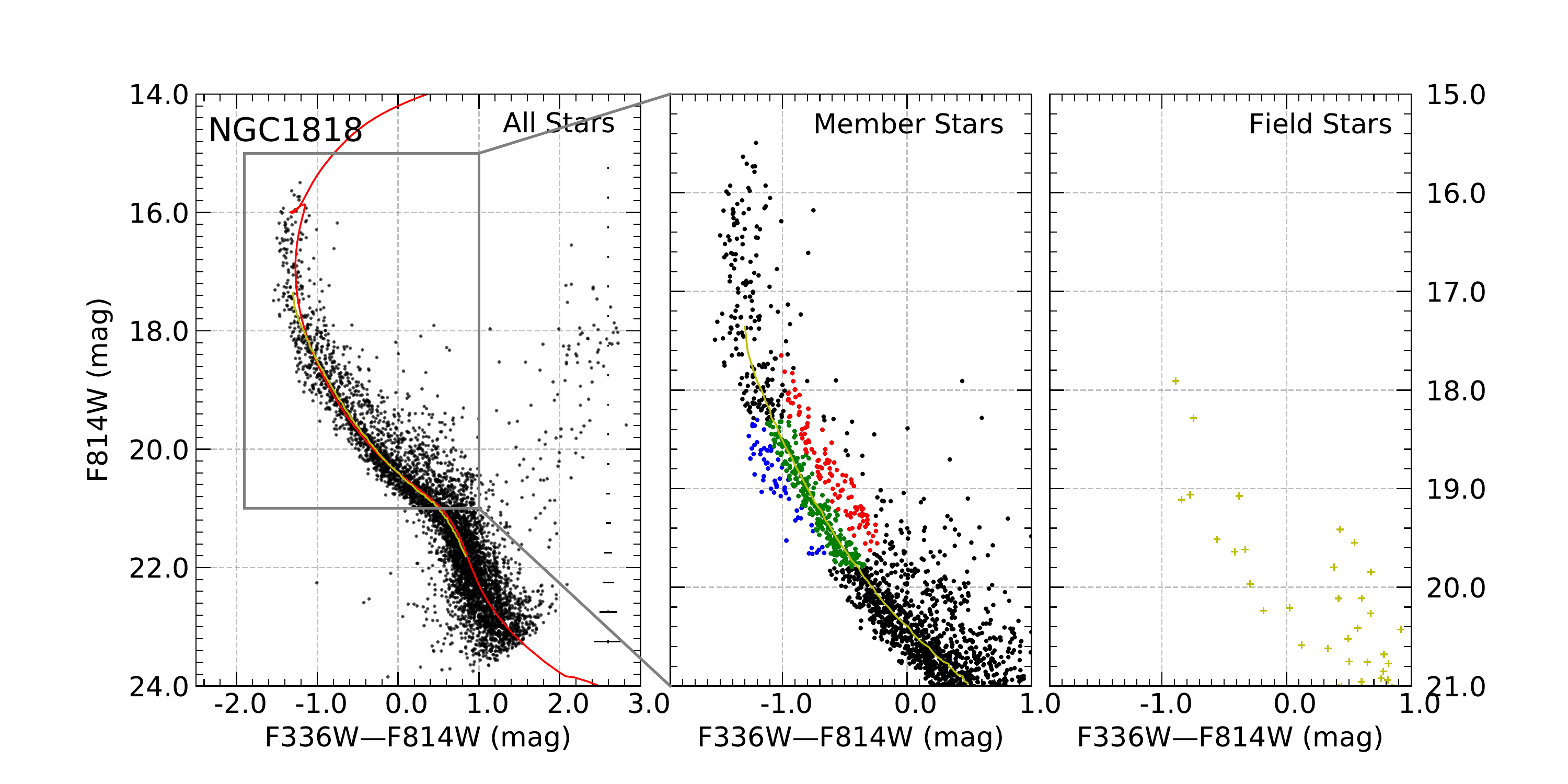}
\caption{CMDs of NGC 1755, NGC 1805, and NGC 1818. (left) All stars
  (black points) in the F814W versus (F336W -- F814W) CMD. The
  best-fitting isochrones and ridge lines are shown as red and yellow
  solid lines, respectively. Error bars on the right of each CMD
  indicate the uncertainties derived from artificial star tests.
  (middle) Zoomed-in CMDs of stars located inside the full cluster
  field and the main bifurcated regions, showing blue-MS stars (blue
  points), red-MS stars (green points), and high-mass-ratio binaries
  (red points). (right) Field stars (yellow pluses) selected from the
  peripheries of the fields of view.}
\label{fig : cmd1}
\end{figure}

\begin{figure}[htb]
\centering
\includegraphics[width = 0.8 \textwidth]{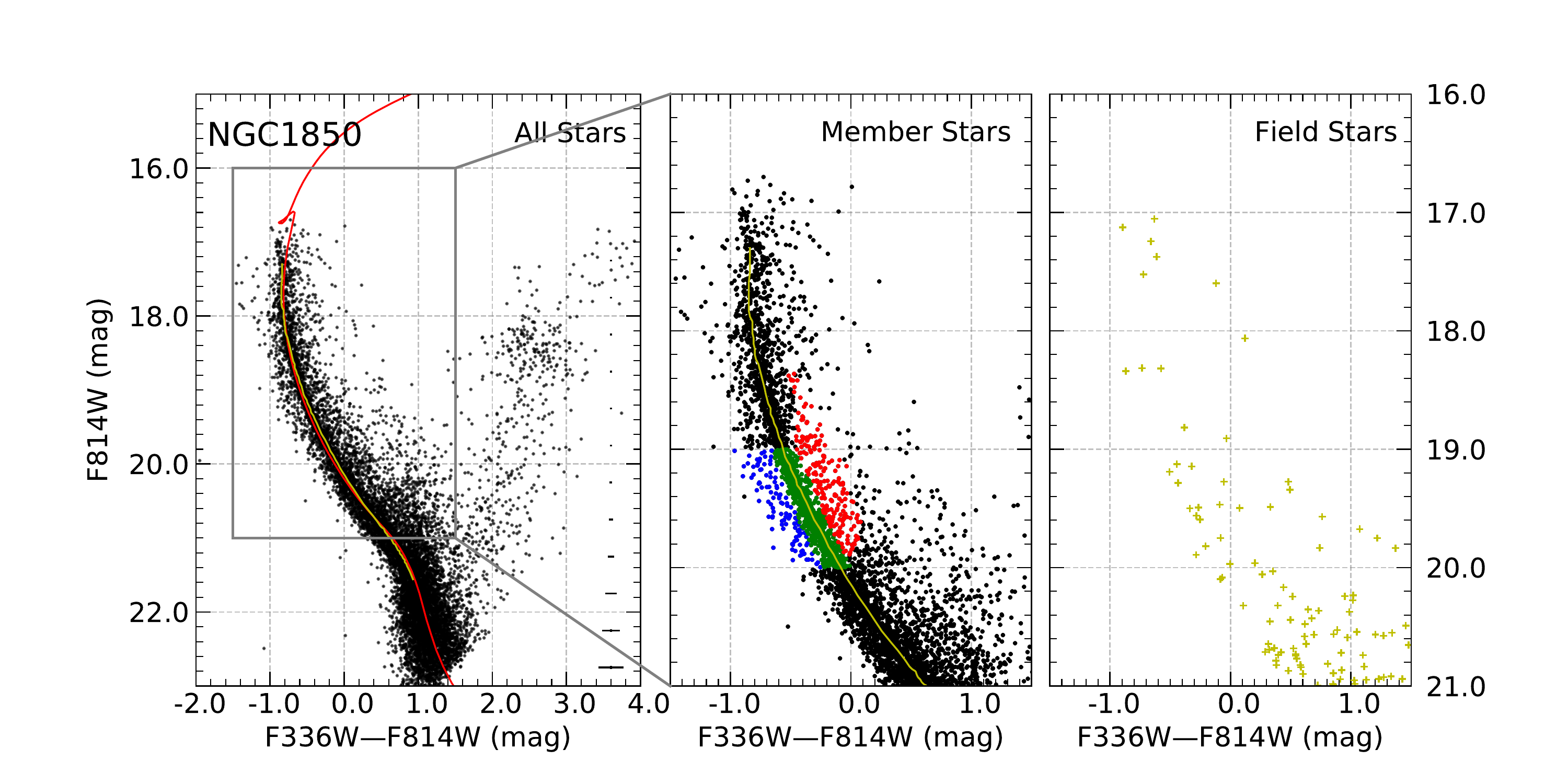}
\vspace{0cm}
\includegraphics[width = 0.8 \textwidth]{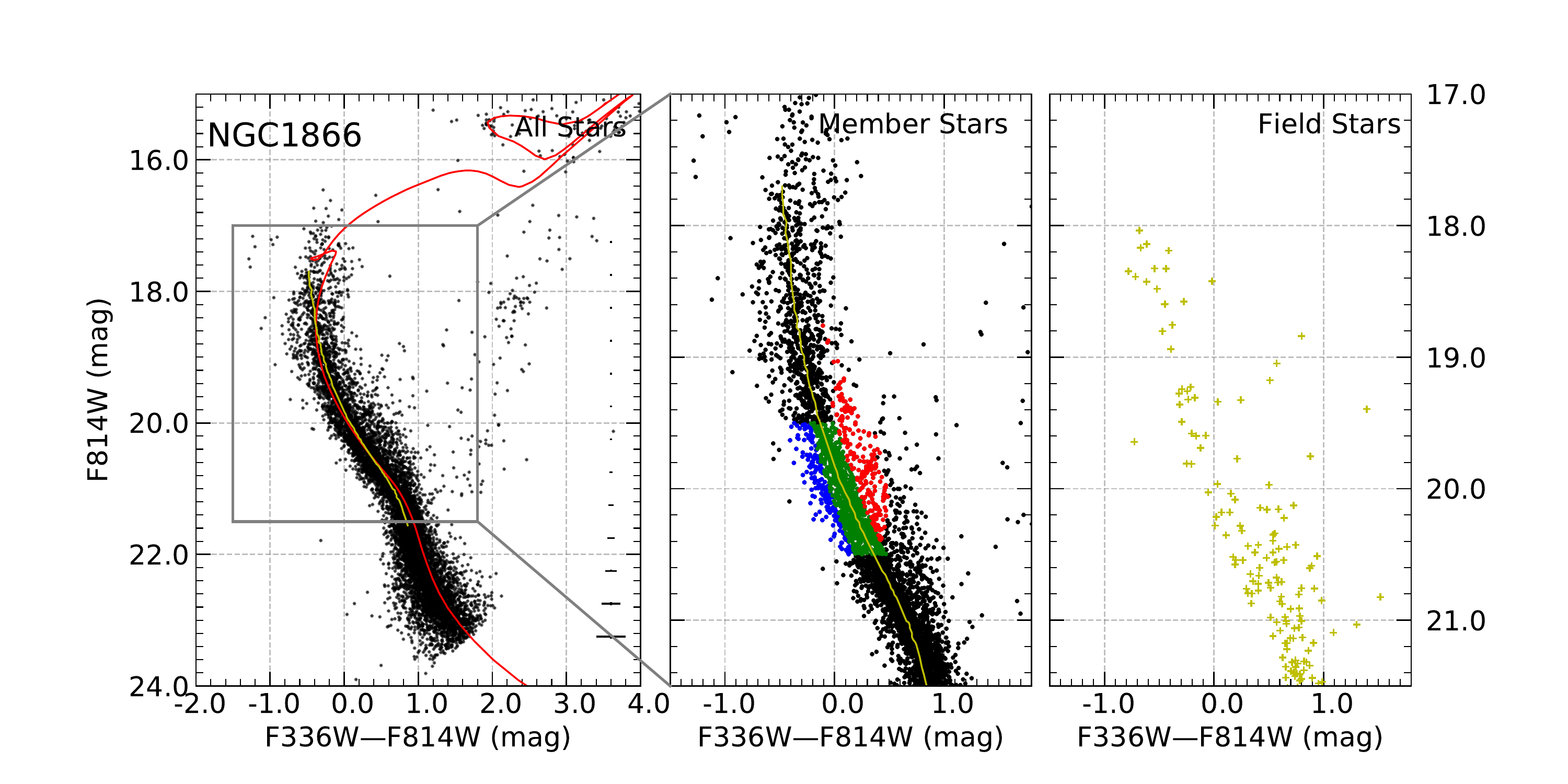}
\vspace{0cm}
\includegraphics[width = 0.8 \textwidth]{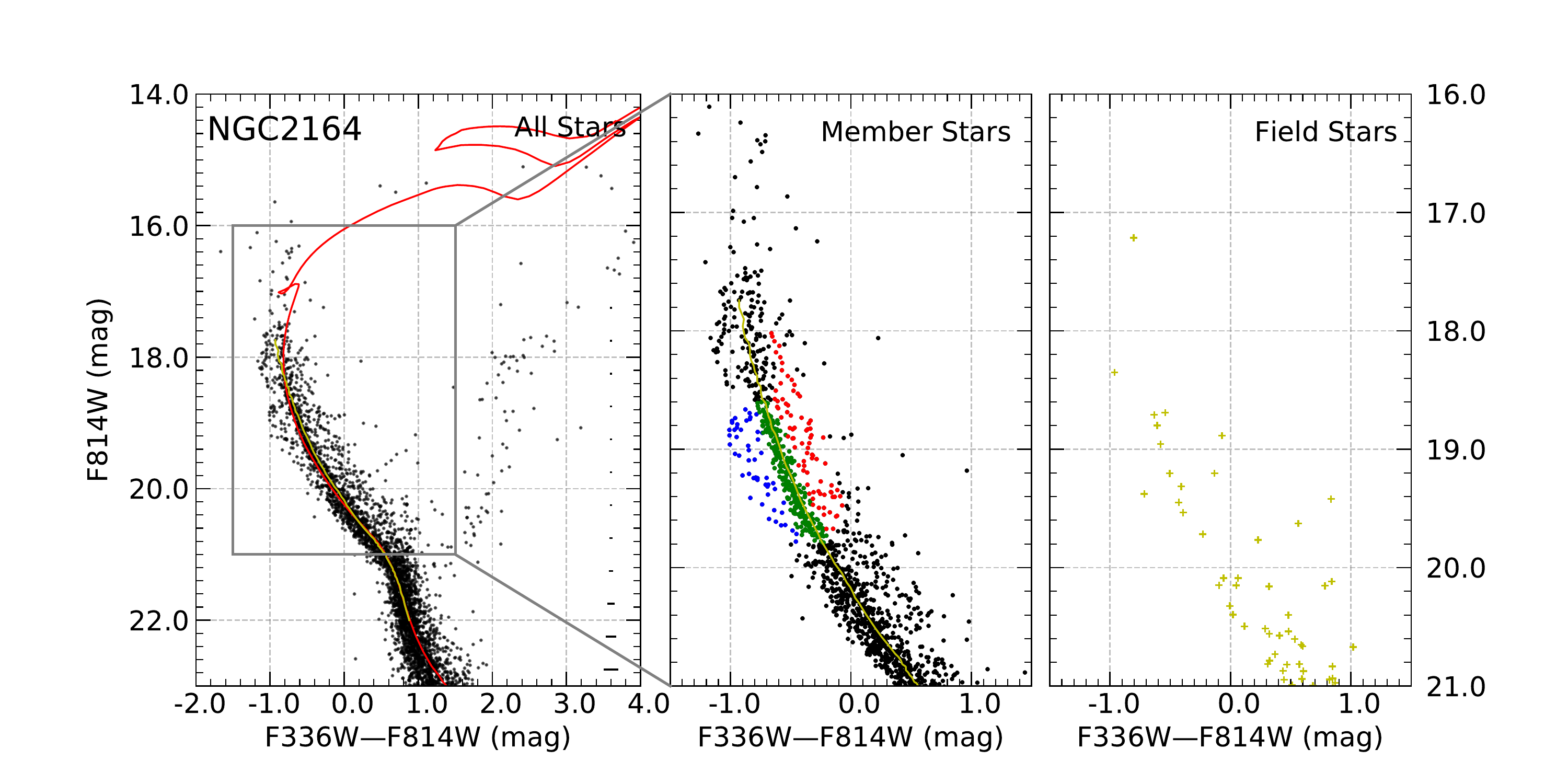}
\caption{As Figure \ref{fig : cmd1} but for NGC 1850, NGC 1866, and
  NGC 2164.}
\label{fig : cmd2}
\end{figure}

\begin{figure}[htb]
\centering
\includegraphics[width = 0.8 \textwidth]{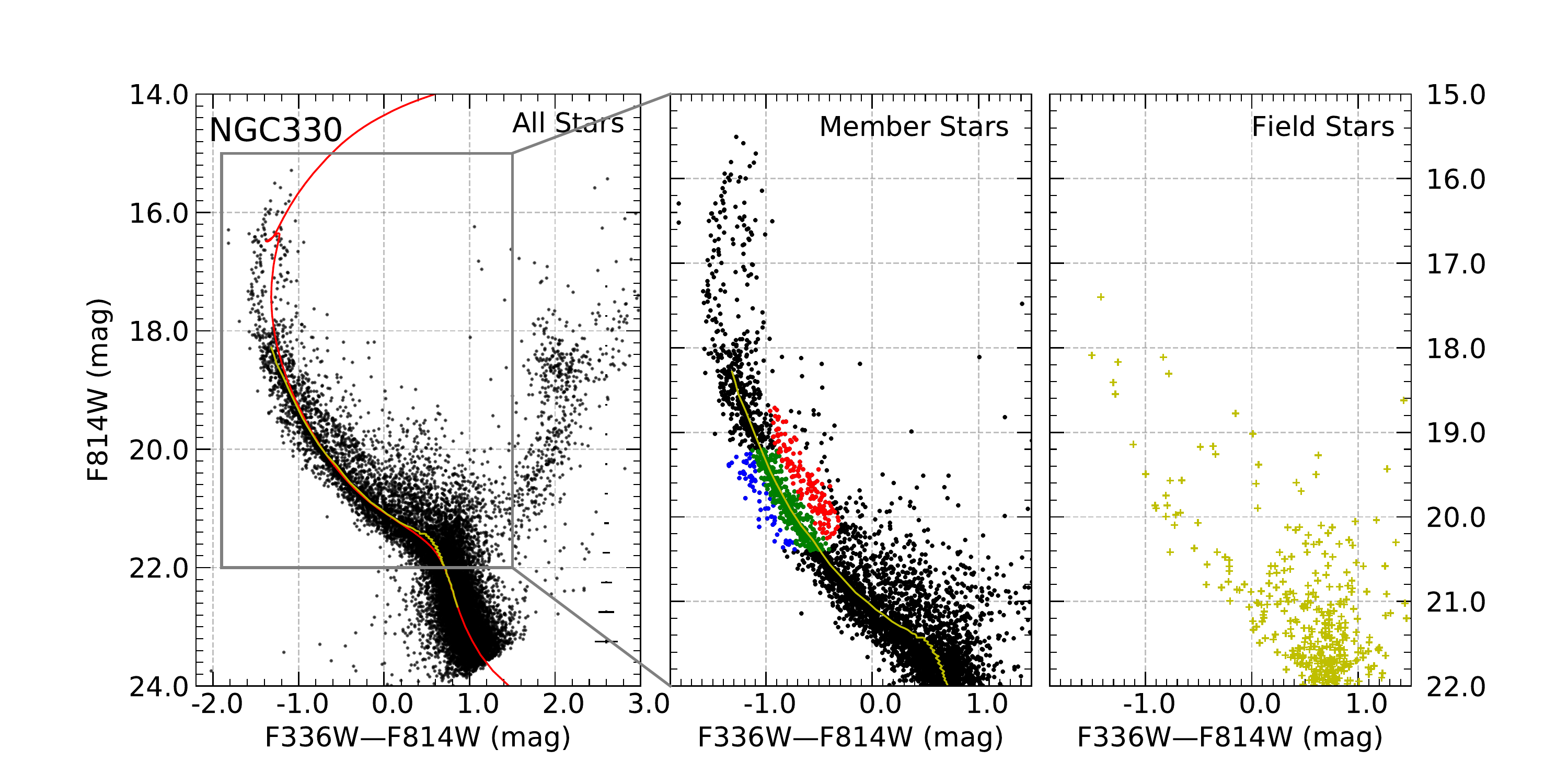}
\caption{As Figure \ref{fig : cmd1} but for NGC 330.}
\label{fig : cmd3}
\end{figure}

\subsection{Differential reddening}

We corrected for differential reddening in the F814W versus
(F336W$-$F814W) CMD. Differential reddening may broaden intrinsically
narrow stellar sequences in the reddening
direction. \cite{milone2012acs} described the detailed method we
adopted to measure and characterize differential reddening. Briefly,
we selected the bulk of the MS stars (whose fiducial line subtends a
wide angle with respect to the reddening direction) as our reference
population. Then, we rotated the original CMD counterclockwise,
adopting the most appropriate angle to align the rotated $x$ axis with
the reddening direction, and measured the horizontal distance between
the reference stars and their fiducial line. These distances represent
the relative reddening suffered by the reference stars compared with
the fiducial sequence. For each star in the catalog, the median value
of the distances of its 40 (spatially) closest reference stars was
assumed as the differential reddening suffered by the target star in
this iteration. An improved rotated CMD could be obtained by
subtracting the median value from the $x$ value of each star. The
newly corrected CMD enabled us to derive a more accurate reference
population and a more precise fiducial line for the next iteration. We
reran this procedure until the new reference population no longer
changed with respect to the previous iteration. The variations in the
reddening direction from all iterations were adopted as the final
differential reddening value. Finally, we rotated the frame back to
obtain the de-reddened CMD.

The relative absorption coefficients at the central wavelengths of the
F336W and F814W filters are $\frac{A_{\rm F336W}}{A_{V}} = 1.65798$
and $\frac{A_{\rm F814W}}{A_{V}} = 0.61018$, respectively, using the
\cite{1989ApJ...345..245C} and \cite{1994ApJ...422..158O} extinction
curves with $R_{V} = 3.1$. Among our star clusters, only NGC 1850
exhibits significant differential reddening: $\vartriangle E({\rm
  F336W}-{\rm F814W})$ ranges from $-0.139 \pm 0.008$ mag to $0.120
\pm 0.017$ mag. The other five clusters do not exhibit significant
variations in $E({\rm F336W}-{\rm F814W})$. As such, the de-reddened
data for NGC 1850 and the original data for the other clusters were
used for further analysis.

\subsection{Isochrone fitting} \label{subsection : isochrone fitting}

To derive the ages, metallicities, distance moduli, and reddening
values for the seven clusters, we followed the method introduced by
\cite{2016ApJ...823...18C}, which assumes that the locus of the bulk
of the stars represents the true underlying isochrone of a given
cluster. First, we derived the fiducial line in the F814W versus
(F336W$-$F814W) CMD based on the 2D probability density function
(PDF), calculated based on kernel density estimation (KDE). The point
maximizing the PDF marks the starting point of the ridge line. We
subsequently traced the ridge line toward both the brighter and
fainter sides by moving along the minimum gradient. A bootstrapping
approach was applied to measure the true fiducial line and the
relevant uncertainties. Specifically, in each round of bootstrapping,
we randomly selected a fixed number of stars (half of the total
number) as a new data set and used them to derive the ridge line using
the method just outlined. After 1000 rounds, an ensemble of ridge-line
points was collected for further processing. The total number of
points was determined by the length of the step used for the KDE PDF
estimation; we adopted 0.01 mag. Then, we distributed all ridge-line
points into 0.05 mag bins. The bin center was assumed to be the final
fiducial magnitude, and half the bin size was assumed as the error on
the magnitude value. The mean value and standard deviation of the
color in each bin were assumed as the fiducial color and uncertainty,
respectively. The bin size adopted should ensure that each bin
contains more than 500 points so as to minimize Poissonian noise.

Second, we compared the fiducial line with a grid of isochrones with
varying, reasonable parameters (ages, distance moduli, and reddening
values). The grid steps adopted for the latter three parameters were 1
Myr, 0.01 mag, and 0.01 mag, respectively. For the cluster
metallicities, we used the typical values, $Z = 0.006$ for the LMC and
$Z = 0.002$ for the Small Magellanic Cloud (SMC). Considering that
these parameters are all distributed uniformly, the best-fitting
isochrone maximizes the likelihood $\mathcal{L}$,
\begin{equation}
\mathcal{L} \simeq {\rm exp} (-\frac{1}{2} \chi^2),
\end{equation}
\begin{equation}
\chi^2 = \sum_{i=1}^N \frac{(\vartriangle {\rm col}_i)^2}{\sigma^2_i},
\end{equation}
where $\vartriangle{\rm col}$ is the difference in color between the
fiducial line and the isochrone at the same magnitude, and $\sigma$ is
the corresponding error. The maximum likelihood corresponds to the
best-fitting isochrone. The uncertainty in each best-fitting parameter
is assumed as the standard deviation of the Gaussian function that
fits the cumulative 1D posterior PDF, which is obtained by
marginalizing the 3D posterior PDF over the other two parameters. The
best-fitting parameters and their uncertainties are listed in
Table~\ref{table : parameter}.

Generally, the distance moduli in Table~\ref{table :
    parameter} are slightly smaller than the mean distances to their
  host galaxies, i.e., $(m-M)_0 = 18.49$ mag for the LMC
  \citep{2014AJ....147..122D} and $(m-M)_0 = 18.96$ mag for the SMC
  \citep{2015AJ....149..179D}. Cluster ages derived in this study are
  slightly older than those derived in the
  literature. \cite{Milone2018} derived their parameters by fitting
  blue- and red-MS stars with non- and rapidly rotating isochrones,
  respectively. The rapidly rotating isochrones yielded ages of 80
  Myr, 50 Myr, 40 Myr, 80 Myr, 200 Myr, 100 Myr, and 40 Myr (following
  the same order as in Table~\ref{table : parameter}). \citet[][their
    Table 6]{Ahumada2018} presented a compilation of ages for NGC
  1755, NGC 1805, and NGC 1818. The age of NGC 1850 inferred from
  other studies is generally found in the range 80--100 Myr
  \citep[e.g.,][]{1993AJ....105..938F,
    2015A&A...575A..62N,2016MNRAS.460L..20B,2017MNRAS.467.3628C,Yang2018}. Ages
  reported for NGC 1866 inferred from isochrone modeling range from
  140 to 220 Myr \citep{Milone2017mnras} and those inferred from
  Cepheids range from 176 to 288 Myr \citep{2019A&A...631A.128C}. Ther
  ages of NGC 2164 and NGC 330 are $\sim$80--100 Myr
  \citep[e.g.,][]{2006ApJ...646..939M, Milone2018} and $\sim$20--50
  Myr \citep[e.g.,][]{ 2000AJ....119.1748K, 2002ApJ...579..275S,
    2005ApJS..161..304M, 2017ApJ...844..119L,
    2020A&A...634A..51B,2020AJ....159..152C}, respectively.

The degeneracies among age, reddening, and distance modulus may
contribute to the large uncertainties associated with those
parameters. We note that these parameters should be used with caution
for two reasons. First, we adopted non-rotating isochrones to fit the
red-MS stars, although they are actually rapidly rotating
stars. Second, the accuracy of parameters derived from isochrone
fitting is reduced when photometric data includes UV
bands. \cite{2018PASP..130c4204B} compared three stellar evolution
models with {\sl HST} photometric data for two globular clusters in
five filters from the UV to the near-infrared. They found that models
yield poor fits to CMDs including a UV filter, which may be due to the
poorly understood effects of reddening at UV wavelengths. Instead of
scrutinizing the specific values of our best-fitting parameters, we
focus instead on the clusters' order, sorted by chronological or
dynamical age.
 
\subsection{Selection of bifurcated MS stars}\label{subsection : region}

\begin{figure}[!htpb]
\begin{center}
\centering
\includegraphics[width=1\textwidth]{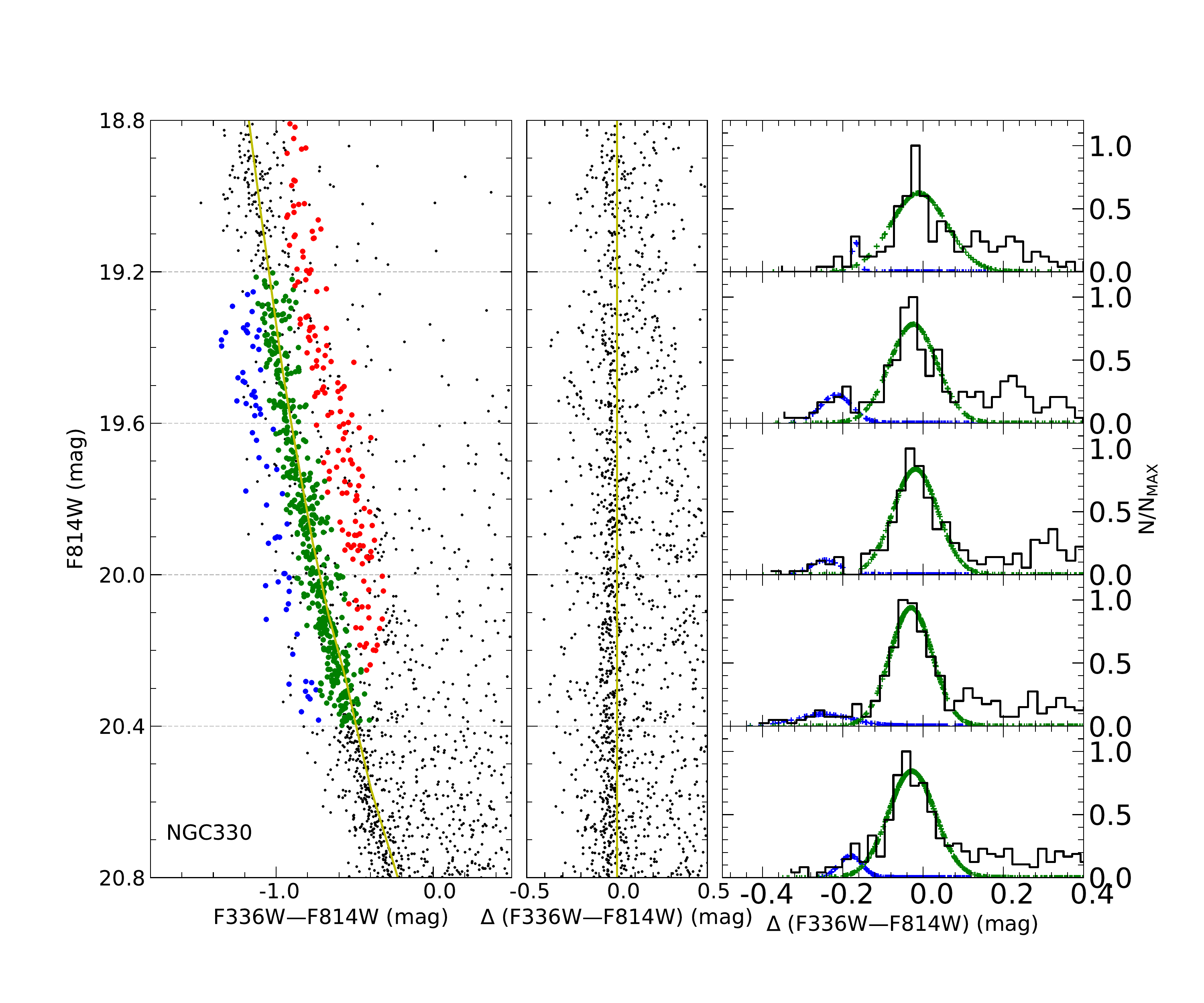}
\caption{(left) Zoomed-in CMD of stars within the cluster area of NGC
  330. The yellow line is the fiducial line of the bulk stellar
  population. Subpopulations selected from Figure~\ref{fig : region}
  are also presented as blue, green, and red points. (middle) MS stars
  in the F814W versus $\vartriangle \rm (F336W - F814W)$ diagram. The
  $x$ axis was obtained by subtracting the fiducial color at the
  corresponding magnitude for each star. (right) Color histograms in
  the five vertically divided bins. Blue and green pluses represent
  the two MS components.}
\end{center}
\label{fig : hist}
\end{figure}

\begin{figure}[!htpb]
\begin{center}
\centering
\includegraphics[width=1\textwidth]{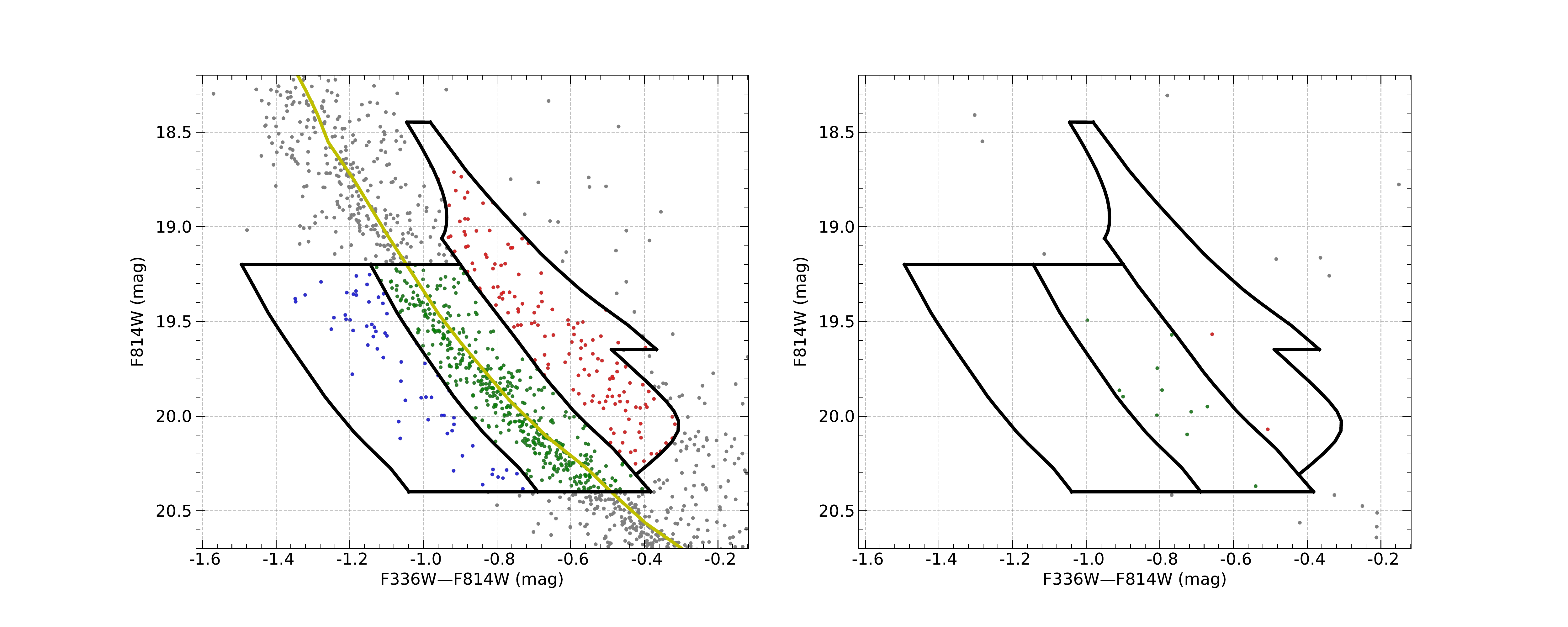}
\caption{Illustration for NGC 330 of our division of MS stars into
  three subpopulations for stars located within the cluster and
  reference fields. The yellow solid line in the left panel is the
  ridge line calculated based on the 2D KDE PDF. Blue-MS stars, red-MS
  stars, and high-mass-ratio binaries are shown as blue, green, and
  red points, respectively. The black frame indicates the boundaries
  governing the subpopulations. The right panel pertains to stars in
  the reference field.}
\end{center}
\label{fig : region}
\end{figure}

We examined the color variations of the MS stars in our cluster sample
and confirmed the presence of bimodal MSs. As an example,
Figure~\ref{fig : hist} (left) shows the zoomed-in CMD of NGC 330, the
verticalized CMD (middle), and the color distributions of MS stars in
five vertically divided bins (right). The color histograms were fitted
using a bi-Gaussian function. The right panel of Figure~\ref{fig :
  hist} shows a clear separation between both components in the middle
bins, which confirms the split-MS phenomenon. The three subpopulations
selected in Figure~\ref{fig : region} are also depicted as blue,
green, and red points in the left panel of Figure~\ref{fig :
  hist}. Their loci are consistent with the main bifurcated region.

The method used to select split-MS stars is similar to that adopted by
\citet{Yang2018}. First, the same photometric procedure was applied to
artificial stars to derive the photometric errors. Artificial stars
were generated with masses following a generic initial mass function
\citep{2001MNRAS.322..231K}, with magnitudes in F336W and F814W
obtained by interpolating the initial masses and magnitudes in the
best-fitting isochrone data table and adopting random spatial
coordinates across the field of view. We added the artificial stars to
the raw images and reran DOLPHOT with the added \textit{Fakestars}
term. To avoid crowding owing to adding artificial stars, we generated
100 stars in each run and collected the input and output catalogs,
repeating this approach more than 1000 times for each chip to minimize
statistical fluctuations. Comparisons between the input and recovered
stars enable us to measure the magnitude uncertainties and also infer
the photometric completeness levels.

\begin{table}[!htb]
\caption{Split-MS regions. (1) Cluster name, (2) Magnitude range, (3)
  Corresponding mass range, (4) Total number of blue-MS stars, (5)
  Total number of red-MS stars, (6) Total number of high-mass-ratio
  binaries, (7) Number of stars within each annular ring.}  \centering
\begin{tabular}{ccccccc}
\hline
\hline
Cluster & F814W range & Mass range       & $N_{\rm bMS}$ & $N_{\rm rMS}$ & $N_{\rm bin}$ &n\\
            &       (mag)        & ($\rm M_\odot$) & &  & &   \\
(1) & (2) & (3) & (4) & (5) & (6) & (7) \\ 
\hline
NGC 1755  & 18.2--19.7 & 2.38--3.68 & 71   & 220 & 79   & 37            \\
NGC 1805  & 18.6--20.1 & 2.10--3.80 & 40   & 149 & 81   & 27            \\
NGC 1818  & 18.3--19.8 & 2.47--4.33 & 59   & 264 & 113 & 43   (49)   \\
NGC 1850  & 19.0--20.0 & 2.35--3.35 & 113 & 963 & 207 & 128 (131) \\
NGC 1866  & 19.5--20.5 & 1.77--2.46 & 198 & 861 & 209 & 126 (134) \\
NGC 2164  & 18.6--19.8 & 2.22--3.20 & 53   & 264 & 80   & 39   (46)    \\
NGC 330   & 19.2--20.4 & 2.35--3.81  & 63   & 467 & 150 & 68              \\
\hline 
\end{tabular}
\label{table : mag range}
\end{table}

Figure~\ref{fig : region} shows the three subpopulations selected from
the bimodal-MS region in NGC 330. We will now first introduce the
theoretical magnitudes of any unresolved binaries, based on the
assumption that no interactions have occurred between both components,
and the entire systems would appear as single, point-like sources in
distant clusters. The magnitude of a binary system is
\begin{equation}
m_{\rm bin} = m_{\rm 1} - 2.5{\rm log}({1+\frac{F_{\rm 2}}{F_{\rm 1}}}),
\label{eq : bin}
\end{equation}
where $m_{\rm 1}$ is the magnitude of the primary star, and $F_{\rm
  1}$ and $F_{\rm 2}$ are the component fluxes. The ratio of the
fluxes is in proportion to the mass ratio. At a given primary mass,
the brightness of the binary system increases with mass ratio and
approaches the brightest point when the two components have identical
masses (0.752 mag brighter than the primary star). The main steps
involved in drawing the black frames in Figure~\ref{fig : region} are
as follows. (1) The ridge line (the yellow line in the left panel)
obtained in Section~\ref{subsection : isochrone fitting} with
3$\sigma$ errors on the blue side was assumed as the blue boundary for
single stars (red-MS stars). (2) Next move the blue boundary
pertaining to the single stars arbitrarily to the left to include most
stars; the enclosed stars are blue-MS stars. (3) The MS--MS binary
sequence with mass ratio $q=0.5$ divides the single stars and
unresolved binaries. Stars residing between the binary sequences at
$q=0.5$ and $q=1$, including the range encompassed by the 3$\sigma$
color errors, are referred to as high-mass-ratio binaries. (4) The
bright and faint cut-offs for MS stars and consistent with the main
bifurcated region were inferred from the histograms in Figure~\ref{fig
  : hist}. The top (bottom) boundary of the high-mass-ratio binary
region contains the trajectory of a series of binaries with increasing
mass ratio combined with a horizontal line with a width equal to the
3$\sigma$ color error at the corresponding magnitudes. For example,
blue-MS stars, red-MS stars, and high-mass-ratio binaries within the
cluster field of NGC 330 are shown as blue, green, and red points,
respectively, in the left panel of Figure~\ref{fig : region}. The
right panel shows the equivalent distributions of stars in the
reference field. A summary of the magnitude bins used to
select split-MS stars is presented in Table ~\ref{table : mag range}, 
which also includes the corresponding stellar masses.

\subsection{Population Ratios}

To study the radial distributions of the selected subpopulations, we
divided them into different radial bins. The population ratios of
blue-MS stars and high-mass-ratio binaries with respect to the full
sample of MS stars are
\begin{equation}
f_{\rm bMS} (r) = \frac{N_{\rm bMS} (r) - A(r)n_{\rm bMS}}{N_{\rm bMS}(r)+N_{\rm rMS}(r)+N_{\rm bin}(r)};
\end{equation}
\begin{equation}
f_{\rm bin} (r) = \frac{N_{\rm bin} (r) - A(r)n_{\rm bin}}{N_{\rm bMS}(r)+N_{\rm rMS}(r)+N_{\rm bin}(r)}.
\end{equation}
where $N_{\rm bMS}(r)$, $N_{\rm rMS}(r)$, and $N_{\rm bin}(r)$ are,
respectively, the numbers of blue-MS stars, red-MS stats, and
high-mass-ratio binaries within radius $r$; $n_{\rm bMS}$, $n_{\rm
  rMS}$, and $n_{\rm bin}$ are the numbers of field stars located in
the same CMD area as the three subpopulations. $A(r)$ is the area
ratio of the cluster field within $r$ and the reference field. All
selected stars were divided equally into annular rings. 
The total number of the three populations and the
 numbers of stars in the annular rings are summarized in Table
  ~\ref{table : mag range}. The values in brackets are the numbers of
  stars in the outermost ring if the mean is not an integer. The
calculated population ratios are shown in Figures~\ref{fig : obs1} and
\ref{fig : obs2}.

The terms $A(r)n_{\rm bMS}$ and $A(r)n_{\rm bin}$ aim to minimize the
effects of field stars. Field-star decontamination was applied under
the assumption that the field-star CMD does not depend on the spatial
coordinates. As such, the CMD of the field stars behind the cluster
region would be similar to that of the reference stars. The area ratio
$A(r)$ was calculated using Monte Carlo simulations. Given the limited
field of the view of the {\sl HST} camera, we selected the images'
margins as our reference fields. Rectangular areas far from the
cluster center were adopted as reference fields, except for NGC 1818
and NGC 1866. Different reference areas were selected for the latter
two clusters, including two field corners for NGC 1818 and the
outermost periphery for NGC 1866.

\section{Results}\label{sec : obsresult}
\begin{figure}[!htpb]
\gridline{\fig{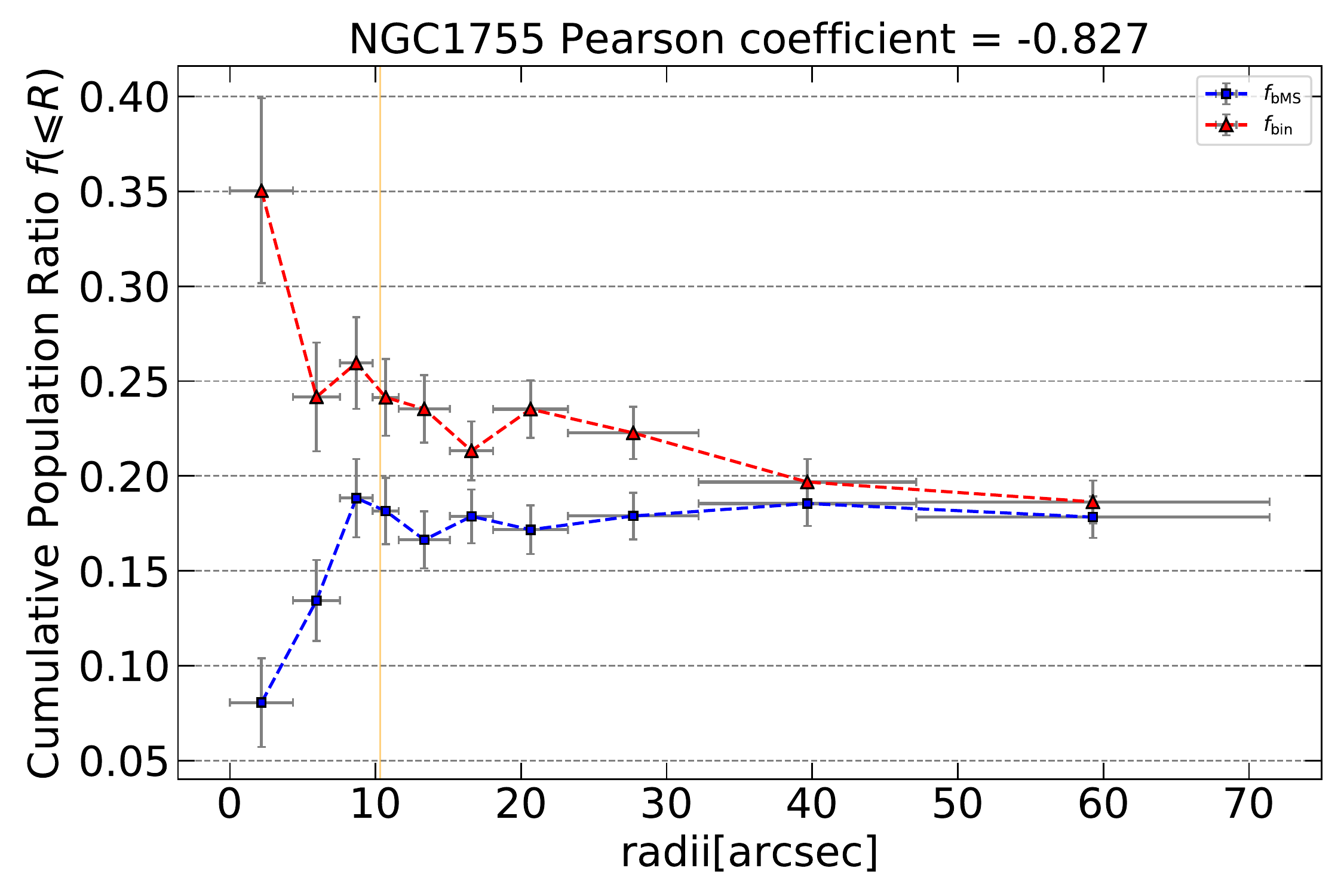} {0.5\textwidth}{}
              \fig{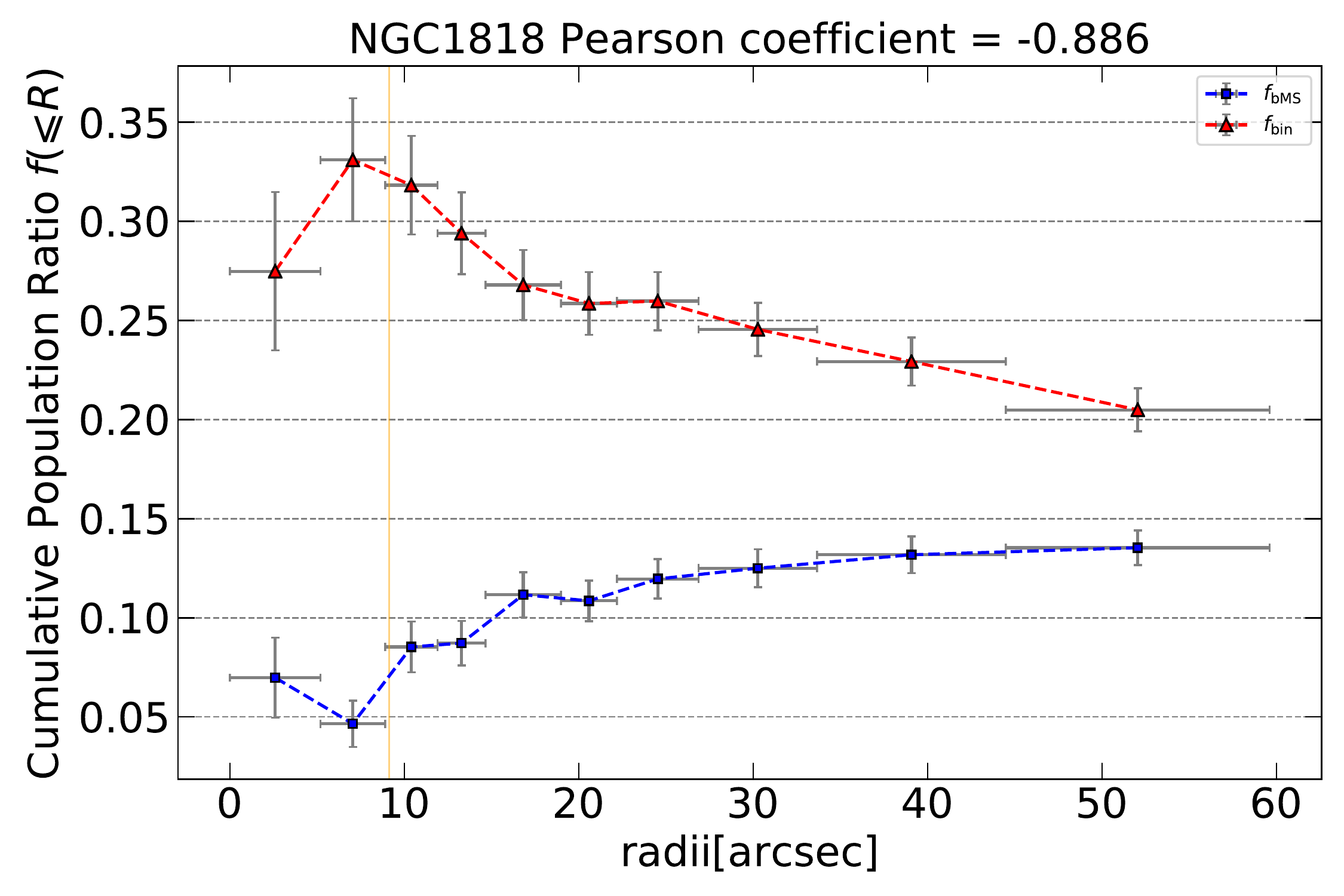} {0.5\textwidth}{}}
\vspace{0cm}              
\gridline{\fig{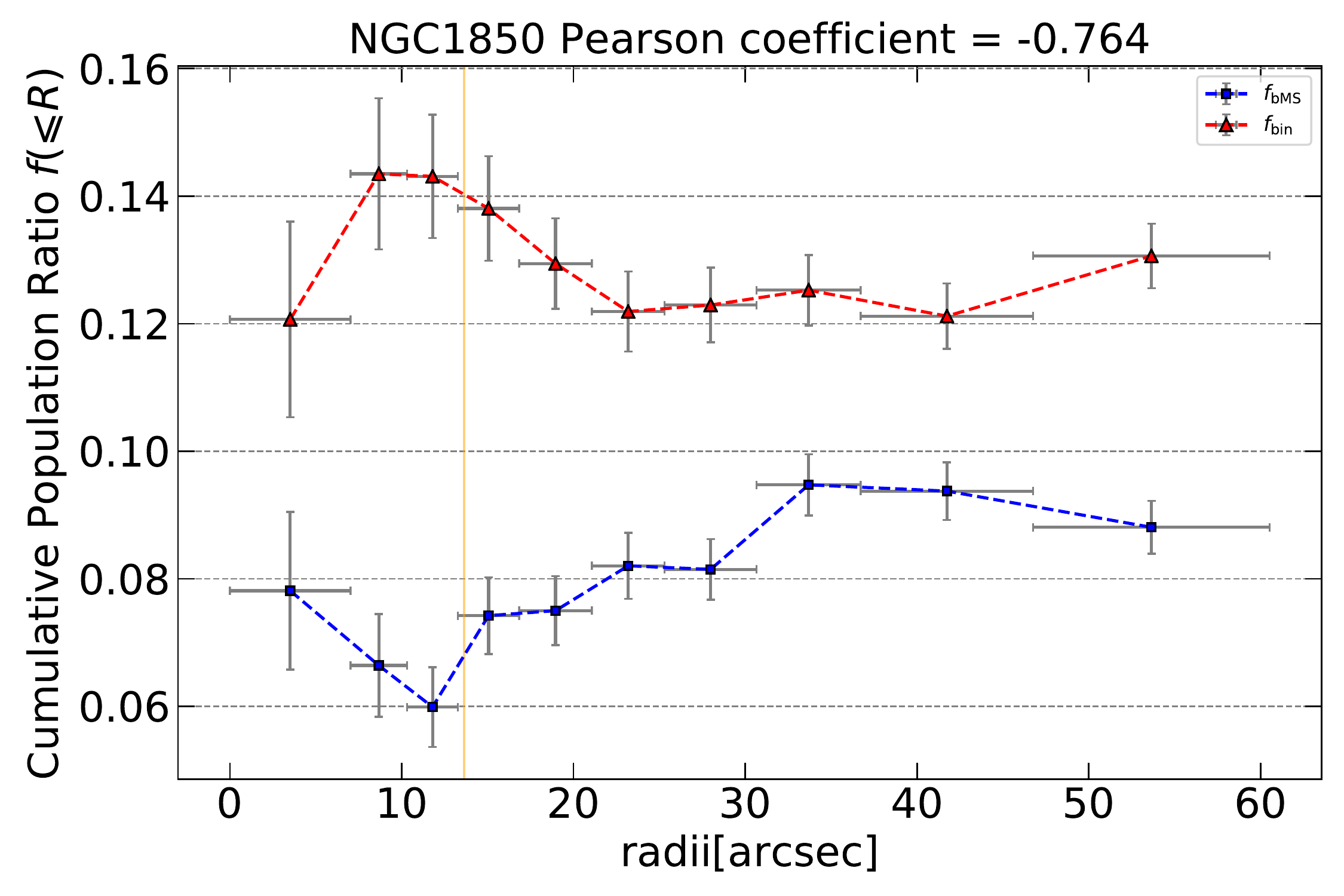} {0.5\textwidth}{}
              \fig{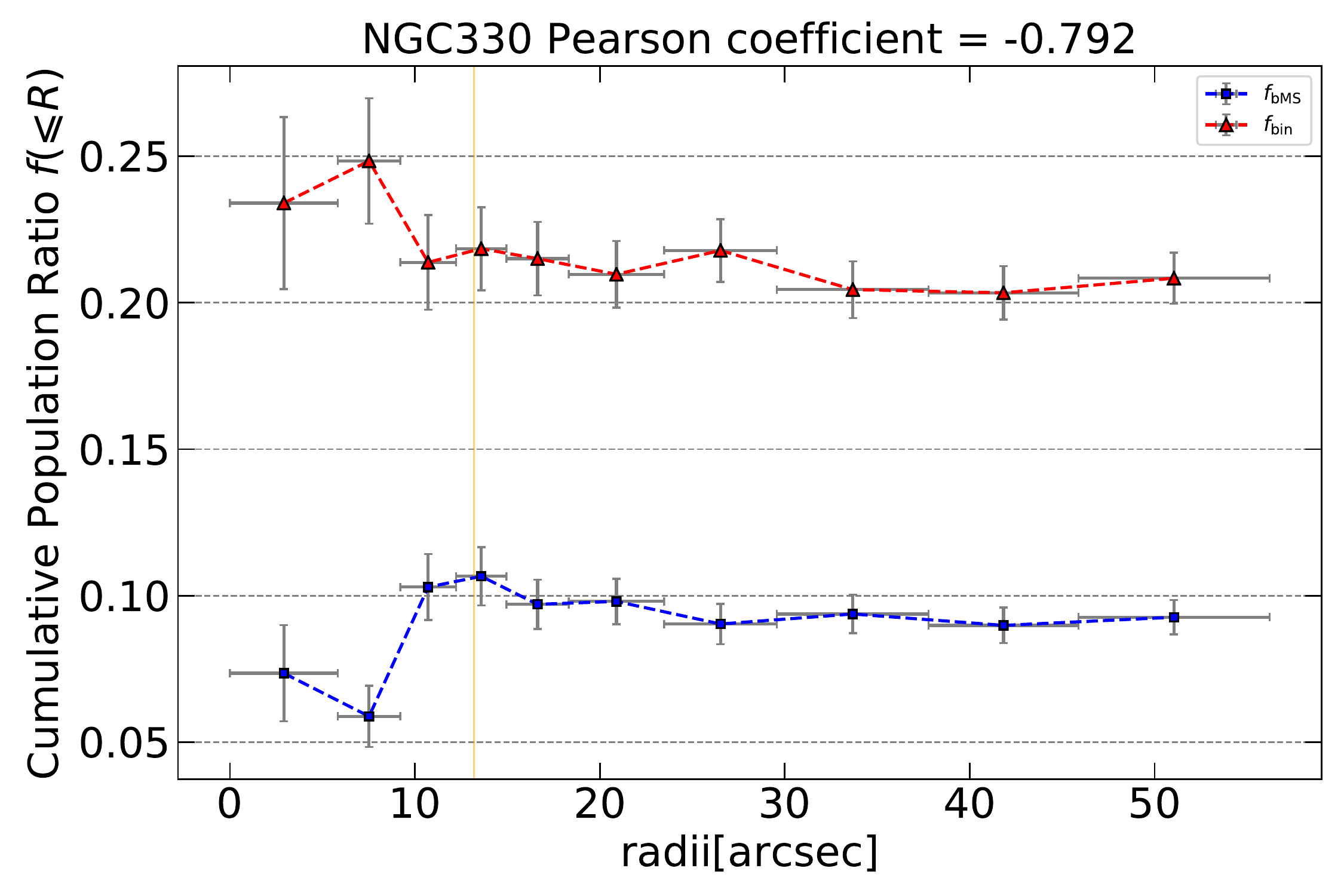} {0.5\textwidth}{}}
\caption{Cumulative population ratios of blue-MS stars (blue squares)
  and high-mass-ratio binaries (red triangles) with respect to the
  full samples of MS stars as a function of radial distance. Cluster
  names (NGC 1755, NGC 1818, NGC 1850, and NGC 330) as well as the
  Pearson correlation coefficients between the two profiles are shown
  in the panel titles. `Error bars' along the $x$ axis represent the
  widths of the annular rings; those along the $y$ axis are Poissonian
  errors. The orange vertical lines indicate the derived half-mass
  radii.}
\label{fig : obs1}
\end{figure}     

\begin{figure}[!htpb]
\gridline{\fig{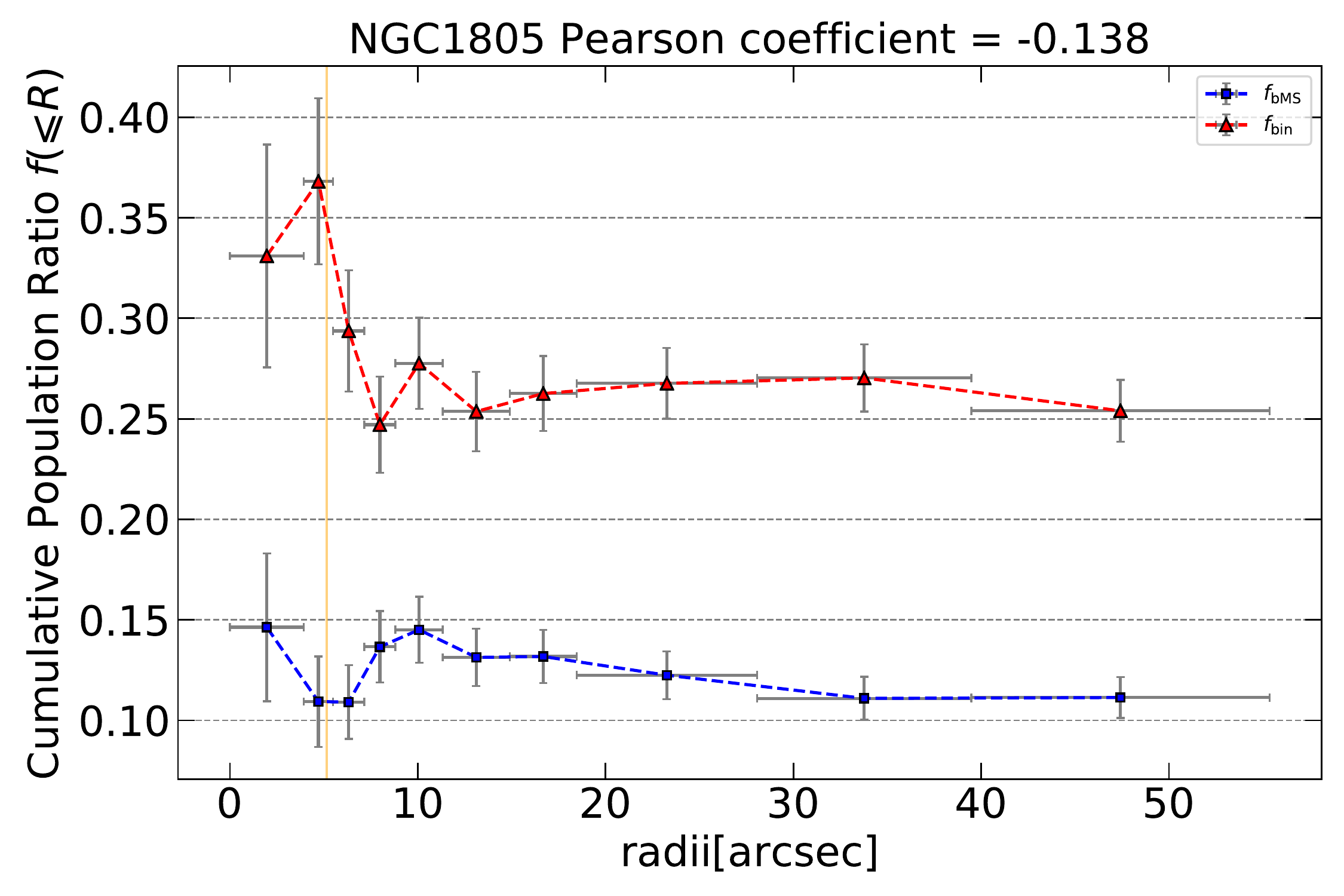} {0.5\textwidth}{}
             \fig{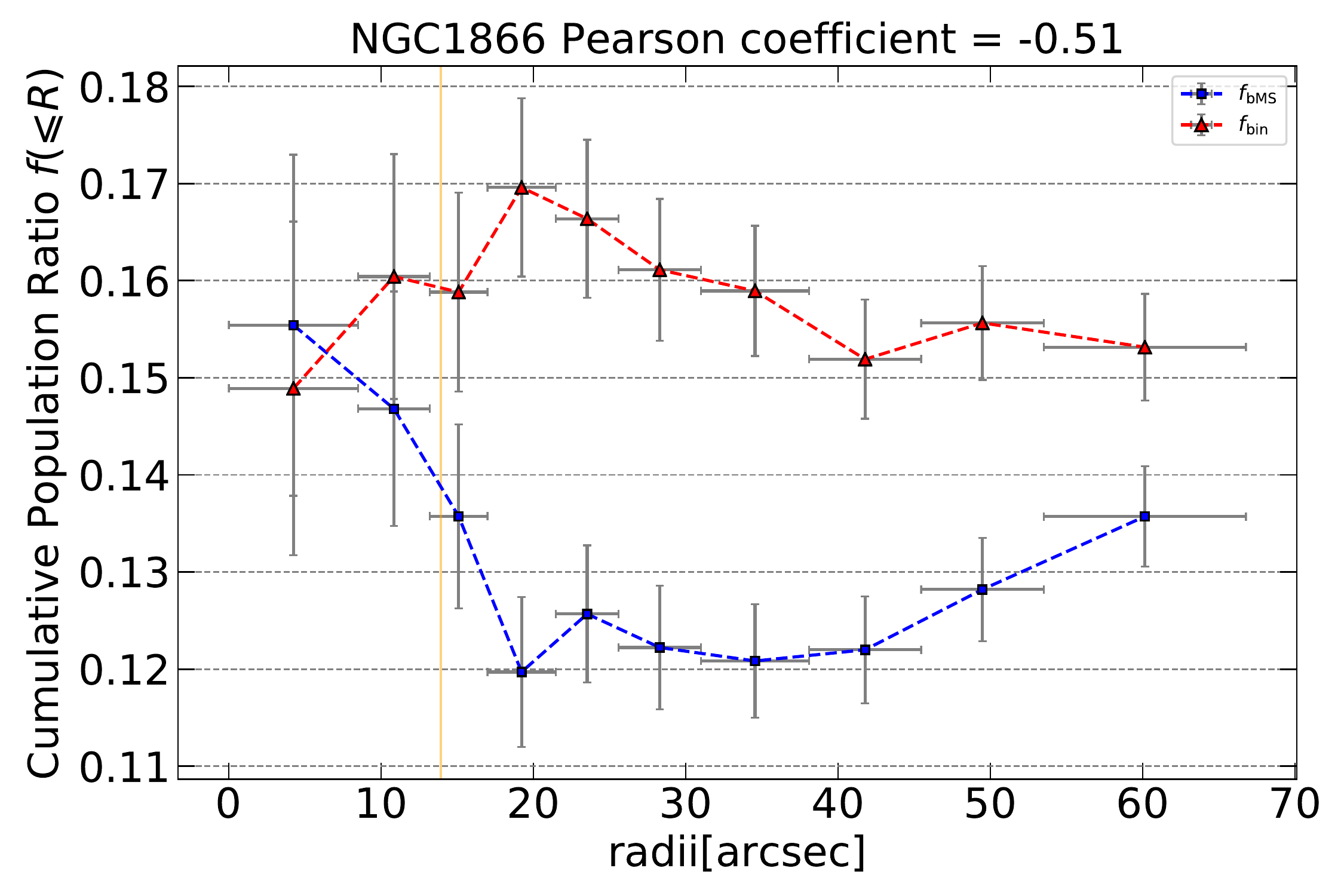} {0.5\textwidth}{}}   
\vspace{0cm}              
\gridline{\fig{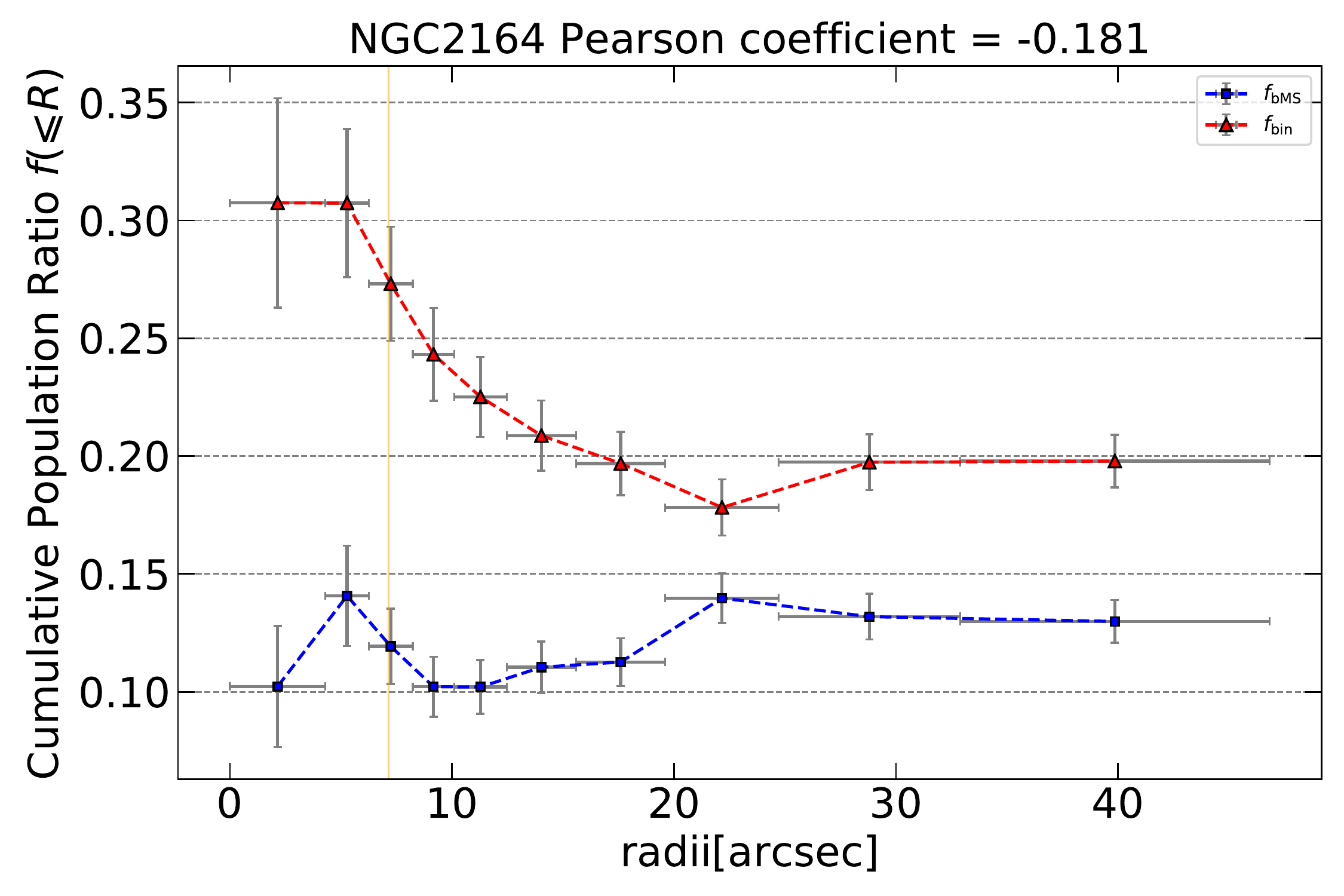} {0.5\textwidth}{}}                 
\caption{As Figure~\ref{fig : obs1}, but for NGC 1805, NGC 1866, and
  NGC 2164.}
\label{fig : obs2}
\end{figure}     

The stellar spatial distribution is a vital ingredient to understand
star cluster formation and evolution. Figures~\ref{fig : obs1} and
\ref{fig : obs2} show the cumulative population ratios of blue-MS
stars and high-mass-ratio binaries as a function of cluster-centric
distance. For each cluster, all bifurcated stars were equally divided
into 10 annular rings. These figures show that the two cumulative
radial population ratios in NGC 1755, NGC 1818, NGC 1850, and NGC 330
(see Figure~\ref{fig : obs1}) are strongly anti-correlated, while no
significant relationship between the population ratios was found in
NGC 1805, NGC 1866, or NGC 2164 (see Figure~\ref{fig : obs2}).

In the first group of clusters, the cumulative population ratio of the
high-mass-ratio binaries generally decreases from the inner region to
the periphery, while the blue-MS stars' profile increases. For
example, in the case of NGC 1755, the red profile drops from 35.0\% to
18.5\%, while the blue profile rises from 8.1\% to 17.7\%, indicating
a higher percentage of high-mass-ratio binaries in the core region and
the gradually increasing importance of blue-MS stars in the outer
regions. Similar profiles were also found for NGC 1818 and NGC 330,
but with smaller amplitudes. It is hard to conclude that the two
populations in NGC 1850 exhibit monotonous trends; their opposite
behavior results in a negative Pearson relation coefficient. Similar
trends for split-MS stars selected based on F438W versus
(F336W$-$F438W) CMD analysis have been observed for NGC 1850 by
\cite{Yang2018}.

In the second group of clusters, no significant correlation was
found. NGC 1805 contains the fewest split-MS stars (27 stars in each
bin), thus contributing large uncertainties to the population
ratios. Adopting a larger bin size, e.g., 50 stars in each annular
ring, would reduce the fluctuations. The top left panel of
Figure~\ref{fig : obs2} shows that the binary profile in NGC 1805
first decreases and then slightly increases at the tail end (large
radii), while the blue-MS profile slightly decreases within the first
three points, resulting in a low negative correlation. In NGC 1866,
the trends in the two profiles result in a medium negative
correlation. Following \cite{Milone2017mnras}, who also studied the
annular population ratios of blue-MS stars and high-mass-ratio
binaries, we calculated the annular population ratios of split-MS
stars selected from the same magnitude ranges and the same radial bin
intervals. The results generally agree with the literature
\cite[see][their Figure 9]{Milone2017mnras}, showing a higher fraction
of blue-MS stars to high-mass-ratio binaries beyond the cluster core
region. The agreement with \cite{Milone2017mnras} confirms the
accuracy of our photometry and data analysis approach. In NGC 2164,
even though the fraction of high-mass-ratio binaries decreases as a
function of cluster-centric distance, no significant anti-correlation
was found between both profiles.

We ran $N$-body simulations to check the spatial distribution of
binary systems in an isolated cluster using the high-performance code
PeTar\footnote{https://github.com/lwang-astro/PeTar}
\citep{2020MNRAS.497..536W}. The initial particle masses,
three-dimensional positions, and velocities were generated by the
updated star cluster initial model generator code
MCLUSTER\footnote{https://github.com/lwang-astro/mcluster}
\citep{2011MNRAS.417.2300K,2019MNRAS.484.1843W}. We generated a total
of 100,000 particles (including 30,000 randomly paired binaries),
following a Plummer density profile with the following parameters:
degree of mass segregation, $S = 0$ (no segregation); fractal
dimension, $D = 3.0$ (no fractality); virial ratio, $Q = 0.5$;
half-mass radius, $R_{\rm h} = 0.5 $ pc; and a
\cite{2001MNRAS.322..231K} initial mass function. The semi-major axis
of the binary population was assigned following
\cite[]{1995MNRAS.277.1491K,1995MNRAS.277.1507K} and
\cite{2012Sci...337..444S} using period distributions for the
binaries' primary components with masses $< 5 M_{\odot}$ and higher
than the mass threshold, respectively. The simulated cluster was
evolved to a maximum age of 300 Myr, yielding output snapshots in time
intervals of 1 Myr. The snapshot data were first processed with the
built-in tools within PeTar to detect binaries and calculate
parameters including the Lagrangian and core radii, average masses,
and velocity dispersions.

Binary systems can be classified as hard or soft by comparing their
binding energy with the surrounding population's kinetic energy. The
boundary between soft and hard binaries is derived as
\begin{equation}
a_{\rm hs} = \frac{GM_{\rm 1}M_{\rm 2}}{3 \langle M_{\rm T} \rangle \sigma_{\rm v, 3D}^2},
\end{equation}
where $\langle M_{\rm T} \rangle$ is the average mass of the
surrounding stars and $\sigma_{\rm v,3D}$ is the three-dimensional
velocity dispersion. Binaries with $a_{\rm hs}$ larger than their
semi-major axis are classified as soft binaries; otherwise, they are
classified as hard binaries. The different spatial distributions of
the two types of binaries can be predicted, since soft binaries become
softer and hard binaries get harder after numerous encounters with
neighboring stars.

\begin{figure}[!tpb]
\begin{center}
\centering
\includegraphics[width=1\textwidth]{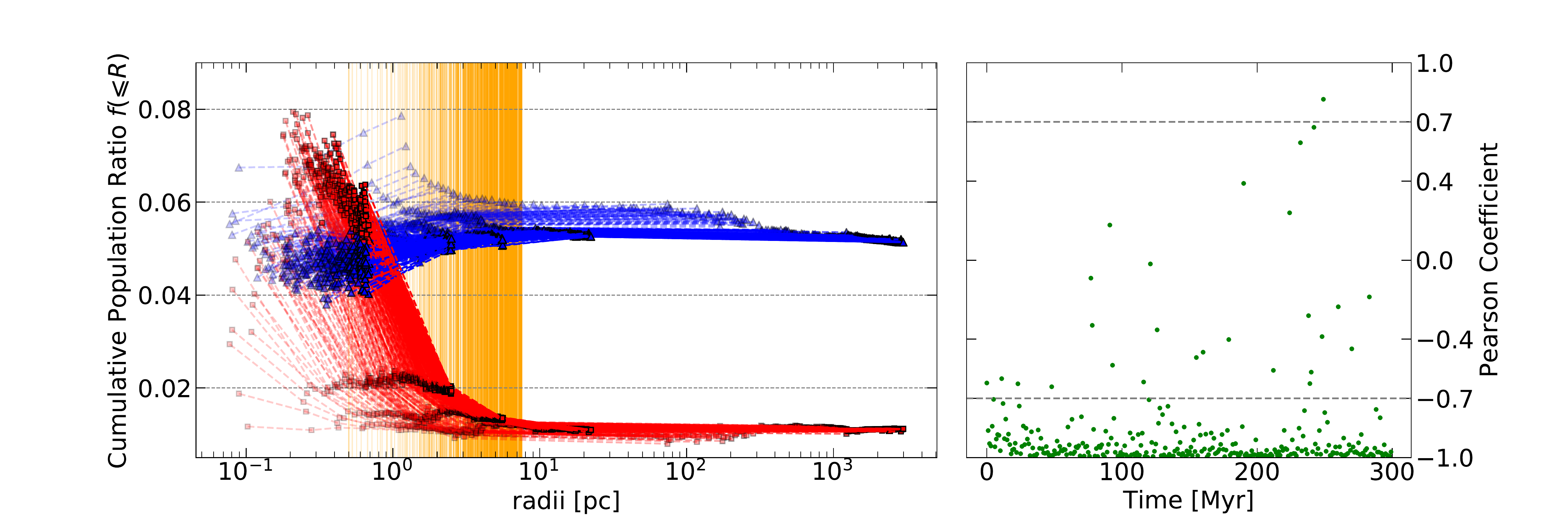}
\caption{$N$-body simulation results. (left) Secular evolution of the
  cumulative population ratios of hard and soft binaries. Based on
  each snapshot, five blue triangles (red squares) indicate the number
  fractions of soft (hard) binaries within the 10\%, 30\%, 50\%, 70\%,
  and 90\% Lagrangian radii. Orange vertical lines indicate the 50\%
  Lagrangian radius, which expands with time. The evolutionary
  timescale is encoded in the transparency of the lines. (right)
  Pearson correlation coefficient between the two profiles as a
  function of evolution timescale. }
\end{center}
\label{fig : simhard}
\end{figure}

Our $N$-body simulation suggests that the anti-correlation found in
the first four clusters may be associated with their binaries'
dynamical evolution. The left and right panels of Figure~\ref{fig :
  simhard} show the cumulative number fractions of, respectively, hard
binaries (red squares) and soft binaries (blue triangles) in all
snapshots and the corresponding Pearson correlation coefficients. For
each snapshot, the number ratio of hard/soft binaries with respect to
the whole sample was measured at five radial distances (10\%, 30\%,
50\%, 70\%, and 90\% Lagrangian radii). The radius of the outermost
bin and the 50\% Lagrangian radius grow with time, indicating cluster
expansion at a young age. In general, soft binaries apparently
dissolve within the first 20 Myr and the dissolution subsequently
slows down due to violent relaxation and two-body relaxation; hard
binaries continuously grow due to the hardness of primordial binaries
and the newly formed dynamical binaries. A high stellar density
reinforces the dissolution or formation of binaries in the core
region, which can be inferred from the significant increase in the
population ratios in the innermost bin. We calculated the population
ratios in each snapshot. The profile of the soft binaries rises as the
cluster radius moves outward, while the profile of hard binaries
declines. The increasing or decreasing trend is enhanced as a function
of time.

The right panel of Figure~\ref{fig : simhard} shows the Pearson
correlation coefficients relating the two profiles (green dots), with
grey dashed lines showing the criteria for strong correlation, i.e.,
an absolute value of the coefficient greater than 0.7. Most of the
green dots have values $< -0.7$, indicating a strong negative
relationship between the spatial distribution profiles of the soft and
hard binaries.

We also checked the radial distributions of the binaries divided by
other parameters, such as orbital period and mass ratio. The
cumulative population ratios of short- and long-period binaries are
negatively correlated, while those of low- and high-mass-ratio
binaries are positively correlated. This can be expected, since most
short-period binaries are hard binaries and binary dissolution has
only a minor dependence on their mass ratios.

Our simulation results provide detailed information about the
distribution of binaries in an isolated cluster, which can help us
constrain the origin of blue-MS stars. In view of a star cluster's
dynamical evolution, primordial binaries go through a rapid
dissolution period (continuing for about 20 Myr) during their violent
relaxation phase. Subsequently, two-body interaction dominates the
dynamics, which reduces the dissolution rate. We analyzed the
population ratios of various stellar populations at an age of 150 Myr
as a concrete example. The fraction of binaries with mass ratio $q
\geq 0.5$ declines from 6.8\% to 4.2\% when the volume expands from
the 10\% to the 90\% Lagrangian radii. This is consistent with the
decreasing radial trend of high-mass-ratio binaries in our
observations. Varying the number of initial binaries can adjust the
final percentage to agree with the observations. The population ratios
of blue-MS stars and simulated soft binaries both increase from the
cluster core to the outskirts, but their percentages differ
significantly. In the simulation, soft binaries occupy almost 80\% of
the entire binary population after the violent dissolution phase, a
much higher percentage than that of binaries with mass ratios $q \geq
0.5$ (which is around 60\%). Constraining the soft binary population
by their mass ratios, periods, or other parameters may lower the
percentage value and retain an increasing cumulative profile. However,
the observational data do not support such approaches. Thus, we
speculate that blue-MS stars are, at least in part, soft binaries, and
their dissolution leads to the observed dip in the cluster center. The
full ingredients of blue-MS stars can only be confirmed once more
observational data become available.

\section{Discussion}\label{sec : dis}

The reason for the different spatial distributions of blue-MS stars
and high-mass-ratio binaries into the two groups of clusters is not
clear. The split-MS stars generally range from 18.2 mag to 20.5 mag,
with slight variations from cluster to cluster. The corresponding mean
masses of red-MS stars in the first group of clusters are generally
larger than those in the second group, indicating the anti-correlated
distributions in the young massive populations. However, NGC 1805 is
an exception. Its split-MS stars are as massive as those in NGC 1818,
but we did not find a strong relationship between its stellar
populations. NGC 1805 and NGC 1818 have similar masses and ages but
different dynamical ages. The dynamical ages are the ratios of the
clusters' ages and their half-mass relaxation timescales. The
half-mass relaxation timescale is given by \citep{Meylan1987a}
\begin{equation}
t_{\rm rh} = (8.92\times 10^5 {\rm yr})\frac{(M/1M_\odot)^{1/2}}{(\bar
  m /1M_\odot)}\frac{(r_{\rm h}/1\, {\rm pc})^{3/2}}{\log(0.4M/ \bar
  m)},
\end{equation}
where $M$ is the total cluster mass, $\bar m$ is the average mass of
its member stars, and $r_{\rm h} $ is the half-mass radius. We derived
dynamical ages for all seven clusters and ordered them from
dynamically young to old--- NGC 1850, NGC 330, NGC 1818, NGC 1866, NGC
1755, NGC 1805, NGC 2164---which is consistent with the order
estimated by \cite{2005ApJS..161..304M} (the latter study did not
include NGC 1755). Thus, we suggest that the anti-correlation shown in
four of our clusters is likely related to both their chronological and
dynamical ages.

Note that the Pearson coefficient may vary for the magnitude
  ranges applied to select bifurcated MS stars. As described in
  Section ~\ref{subsection : region}, the magnitude cuts were visually
  inferred from the color histograms of Figure ~\ref{fig : hist}. To
  check the influence of our visual inspection, we compared the
  spatial distributions of stars within the same mass range (2.5--3.2
  $M_\odot$) in all clusters (NGC 1866 was not included since its
  average mass is much smaller than the other cluster masses). The
  results showed a medium negative correlation in NGC 1850 and NGC
  330. The reduced number of analyzed stars may contribute to this
  variation. Nevertheless, a shortage of blue-MS stars in the centers
  of those clusters was also observed, thus supporting our main
  conclusion that blue-MS stars may be partially associated with soft
  binaries. Considering the large uncertainties in stellar masses
  inferred from the best-fitting isochrones and the small numbers of
  bifurcated stars, we only applied magnitude cuts.

Does the observed spatial distribution originate from the clusters'
initial conditions? We applied Kolmogorov--Smirnov (K--S) tests to
quantify the influence of the initial conditions. As dynamic processes
are more intensive in the center than in the periphery, the outer
regions may still retain traces of the initial distribution. Comparing
the bifurcated MS stars' radial behavior within the half-mass radius
and across the full cluster field may reveal signatures of the initial
distribution. For high-mass-ratio binaries, K--S tests in the inner
region generally yield $p$ values greater than 0.5, indicating that
they have similar radial distributions as the red-MS population. The
similarity decreases when we enlarge the area to encompass the entire
cluster field; the $p$ values in some clusters are low enough to
reject the null hypothesis that the two population were drawn from the
same distribution (e.g., $p = 0.06$ for NGC 2164). For blue-MS stars,
the $p$ values pertaining to NGC 1818 and NGC 330 within the half-mass
radius are lower than the significance level of 0.05 and become larger
when K--S tests are applied to populations within the entire cluster
field; the opposite applies to NGC 1850 and NGC 1866, i.e., the $p$
values within the entire cluster field are $< 0.05$. K--S tests
applied to the blue- and red-MS stars in the other three clusters
yield $p$-values larger than 0.1 at different radii.  The blue-MS
stars' small numbers may contribute to the uncertainties, since the
K--S test's statistical power increases with sample size. Significant
contamination by field stars in the outer regions also affects the
K--S test's accuracy. In summary, we cannot conclude whether the
observed radial distributions are the result of dynamical processes or
retained from the initial conditions.

The dynamical stage of a star cluster can also be reflected by a
population's cluster-centric concentration. The parameter $A^+$ was
used to evaluate the concentration of stars. It was introduced by
\cite{Alessandrini2016} to measure the central concentration of blue
straggler stars. The value of $A^+$ represents the area enclosed by
the cumulative distributions of the population of interest and the
reference population,
\begin{equation}
A^+ (x) = \int ^x _{x_{\rm min}} \phi_{\rm BSS} (x') - \phi_{\rm REF}(x') {\rm d}x',
\label{eq : A}
\end{equation}
where $x = \log (r/r_{\rm h})$ is the logarithm of the cluster-centric
distance normalized to the half-mass radius, $r_{\rm h}$, and $x_{\rm
  min}$ is the minimum value. We calculated the $A^+$ values of the
populations in Figures~\ref{fig : obs1} and \ref{fig : obs2},
selecting red-MS stars within the same spatial area as the reference
population. In NGC 1755 and NGC 1818, high-mass-ratio binaries have
positive $A^+$ values, and blue-MS stars have negative $A^+$ values in
all radial bins. This clear separation indicates that high-mass-ratio
binaries are more concentrated than blue-MS stars. The binaries'
higher $A^+$ values were also observed in NGC 1850 and NGC 2164,
although their values overlapped somewhat with the $A^+$ of blue-MS
stars. In NGC 1805, NGC 1866, and NGC 330, no significant differences
were spotted between the populations' $A^+$ values. All clusters in
Figure~\ref{fig : obs1}, except NGC 330, exhibit a strong
concentration of high-mass-ratio binaries. In two-body encounters, the
more massive star loses energy and moves inward, while the less
massive star gains energy and moves outward. Cumulative two-body
interactions lead to mass segregation, where more massive stars sink
to the cluster center, while less massive stars predominantly populate
the outer regions of the cluster. The segregation timescale of a star
is inversely proportional to its stellar mass, i.e., massive stars
need a shorter time to sink to the center. Therefore, it can be
expected that $A^+$ of the massive population increases with a
cluster's dynamical age, and studies show that $A^+$ is a good
dynamical indicator for globular clusters. However, the effect of the
initial distribution on $A^+$ is non-negligible. In addition, the
small mass differences between the population of interest and the
reference population and the different tidal forces exerted by their
host galaxies also affect the accuracy of $A^+$ as dynamical indicator
in our young cluster sample.

Their spatial distributions offer a new perspective to analyze the
origin of the blue-MS stars. In recent years, direct measurements of
projected stellar rotation rates in young clusters have revealed
bimodal distributions, which is consistent with the observed color
bimodality in the split-MS and eMSTO regions
\citep{2019ApJ...876..113S, 2020MNRAS.492.2177K}. Although stellar
rotation is now widely accepted as the dominant mechanism behind those
features, single stars alone, even with different rotation rates,
cannot fully explain the observations. In the framework of the
rotation-spread scenario, the blue- and red-MS stars would have
similar spatial distributions since there is no significant mass
difference between single slow- and fast-rotators. The K--S tests
applied to the two populations and the concentration measurements in
this paper are indicative of the different spatial distributions.

\cite{D'Antona2017nature} showed that varying the rotation rates alone
is not sufficient to cover the upper blue-MS. To explain the large
number of slowly rotating massive stars, they linked binary
interactions to the rotation-spread
scenario. \cite{D'Antona2017nature} suggested that the slow rotators
may initially have been rapidly rotating stars that subject to recent
braking. Interior material mixing induced by rotation at early times
makes them look younger than their non-braked counterparts. The
properties of the younger, non-rotating population are in good
agreement with those of the upper blue-MS stars. If tidal torques
cause the braking, the upper blue-MS stars would actually be binary
stars. \cite{2019ApJ...876..113S} also suggested that slow rotators in
the open cluster NGC 2287 may have been slowed down by their binary
components. Efficient interactions require small separations between
the two components, which initially suggest the blue-MS stars are, at
least in part, hard binaries.  However, the observed spatial
distribution suggests a conclusion to the contrary. Our simulation
shows that the central concentration of the hard binaries and the
radially decreasing profile of the cumulative population ratios is
enhanced with time. The radial profile of the soft binaries is more
consistent with that of the observed blue-MS stars, leading us to
conclude that blue-MS stars are associated with soft binaries. It
therefore appears that our observations add more questions about the
origin of the blue-MS stars.
  
We note that velocity dispersion information may be used to
disentangle these binary scenarios, since short-period binaries would
have higher radial-velocity dispersions. We used the measured
heliocentric radial velocities of 14 blue-MS stars and 16 red-MS stars
in NGC 1818 \citep{2018AJ....156..116M}; their radial-velocity
dispersions were 9.50 km s$^{-1}$ and 5.76 km s$^{-1}$,
respectively. However, given the prevailing uncertainties, there are
no significant differences between these values.

The external environment---such as tidal forces and binarity---also
affects a cluster's dynamical evolution and may change the stellar
radial distribution. Clusters experience different tidal forcing due
to their host galaxies, resulting in distinctive spatial
behavior. Studies \citep[e.g.,][]{2019MNRAS.489.4367P,
  2018MNRAS.478.2164P, 2021arXiv210103157P} show that differential
tidal effects lead to variations in the structural parameters and the
internal dynamical evolution of stars in the Milky Way and MC
clusters. In this paper, NGC 1850 is located in the LMC's bar
structure. It is subject to stronger tidal forces than the other
clusters. Moreover, the MCs host a large number of binary
clusters. \citet{Bhatia1991} showed that NGC 1850 is a member of a
cluster pair. \citet{Bica1999} also identified NGC 1755 and NGC 1818
as members of binary clusters and NGC 1850 as a member of a triple
cluster. The tidal force between a cluster pair or among
multiple-cluster systems can accelerate their dynamical evolution,
especially if the components have comparable masses. Although
simulating the tidal force of each cluster is beyond this paper's
scope, we should nevertheless take the external environment into
consideration when we study the spatial distribution of stars in
clusters.

\begin{figure}[!htb]
\begin{center}
\includegraphics[width=1\textwidth]{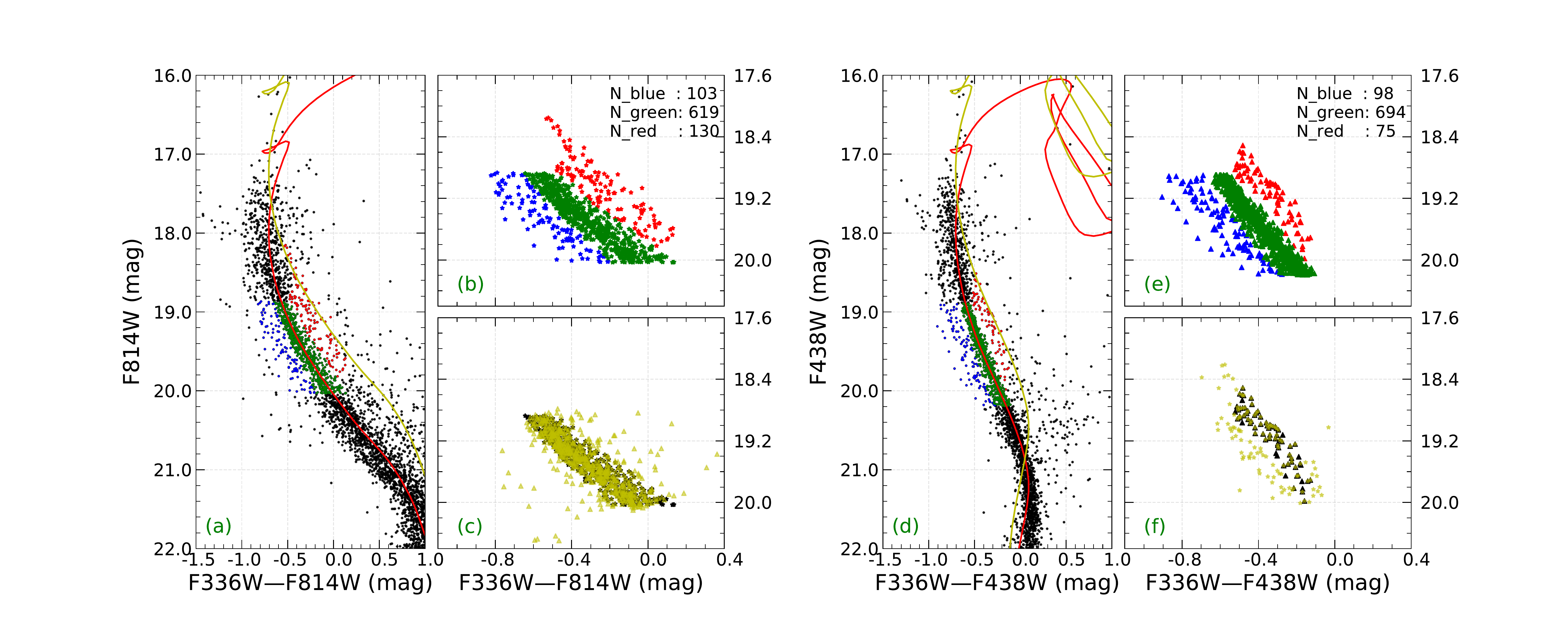}
\caption{Discrepancies between subpopulations selected from two
  different NGC 1850 CMDs. The left panels show stars in the F814W
  versus (F336W$-$F814W) CMD, and the right panels show stars in the
  F438W versus (F336W$-$F438W) CMD. Black points in panels (a) and (d)
  show all stars in the cluster region, with stars classified as
  blue-MS stars, red-MS stars, and high-mass-ratio binaries in the
  main split region shown as blue, green, and red points,
  respectively. The red (yellow) solid line indicates the best-fitting
  isochrone (equal-mass binary sequence). Panels (b) and (e) show a
  zoomed-in version of the three populations, with individual numbers
  shown in the top-right corners. Panels (c) and (f) show comparisons
  of red-MS stars and high-mass-ratio binaries selected from both
  CMDs. Black symbols indicate stars selected from the same frame and
  yellow symbols indicate stars selected from different frames.}
\end{center}
\label{fig : InOb}
\end{figure}    

We note the discrepancies in the NGC 1850 population ratios between
stars selected from the F814W versus (F336W$-$F814W) CMD (this paper)
and the F438W versus (F336W$-$F438W) CMD \citep{Yang2018}. On the one
hand, the two profiles in this paper show more fluctuations, and the
decrease or increase is not as strong as that presented in
\cite{Yang2018}. On the other hand, the number of high-mass-ratio
binaries is always larger than that of the blue-MS stars. To fully
understand the properties of the stellar populations in different
CMDs, we collected raw images observed through the F336W and F438W
filters from program GO-14069 (PI: N. Bastian) and images in F275W and
F814W from program GO-14174 (PI: P. Goudfrooij) using UVIS/WFC3 on
board the {\sl HST}. The post-pipeline processing, photometric
approach, and data selection criteria were identical to the method
introduced in Section~\ref{sec : data}. The final catalog contains
stellar magnitudes in four filters.

Differences in the two sets of subpopulations---selected from the
F814W versus (F336W$-$F814W) and the F438W versus (F336W$-$F438W)
CMDs---are shown in Figure~\ref{fig : InOb}. Panels (a) and (d) show
the distributions of all stars in the two CMDs and highlight the main
split regions; zoomed-in versions are shown in panels (b) and (e). We
note the different numbers of stellar populations in the main split
region, particularly for high-mass-ratio binaries. The number of
binaries (130) selected from panel (b) is nearly twice larger than the
number of binaries (75) in panel (e). Panel (f) shows binaries in the
UV--infrared CMD (yellow asterisks) that overlap with binaries in the
UV--optical CMD (black triangles). It is clear that the yellow
asterisks cover a wider area than the black triangles; see also the
inconsistency with the red-MS stars in panel (c). Therefore, some
red-MS stars may instead be classified as blue-MS stars or
high-mass-ratio binaries.  The smaller number of high-mass-ratio
binaries of \cite{Yang2018} led to the two profiles to cross.

The reason for the increased scatter is not yet clear. One possible
explanation is found in interactions between binary stars. The
theoretical MS--MS binary sequence was calculated by adding the
luminosities of two individual stars, as in Equation~\ref{eq :
  bin}. However, the effects imposed by companions may not follow this
equation, particularly in the case of close binary systems. This
inconsistency in stellar positions in different CMDs may indicate that
the spectral energy distributions of some stars are not as simple as
the theoretical predictions. A filter combination with a longer color
baseline is recommended to select stars in CMDs.
 
\section{Summary}\label{sec : sum}

We derived high-precision photometry from {\sl HST} images of seven MC
clusters to study the spatial distributions of their blue-MS
stars. Stars located in the split-MS region were classified into three
populations, including blue-MS stars, red-MS stars, and
high-mass-ratio binaries. The cumulative population ratios in 10
radial bins out to the clusters' outer regions were calculated. The
results show that the population ratios of high-mass-ratio binaries to
blue-MS stars in four of our clusters (NGC 1755, NGC 1818, NGC 1850,
and NGC 330; see Figure~\ref{fig : obs1}) are strongly
anti-correlated, while no prominent relationship was found for our
other three sample clusters (NGC 1805, NGC 1818, and NGC 2164; see
Figure~\ref{fig : obs2}).

Generally, the number fraction of high-mass-ratio binaries with
respect to the total number of MS stars analyzed decreases as a
function of the radial distance from the cluster center, while the
fraction of blue-MS stars increases. We analyzed the mass range of the
bifurcation in each cluster, and the dynamical stage of the cluster
based on their dynamical age ${t_{\rm iso}}/{t_{\rm rh}}$ and the
dynamical indicator $A^+$. The clusters in Figure~\ref{fig : obs1} are
young, containing more massive split-MS stars. At the same time, they
are also dynamically young. An $N$-body simulation of 100,000
particles, with 30\% binary systems and a Plummer density profile, was
run. It showed that the dissolution of soft binaries dominates a
cluster's dynamical evolution in the first $10^8$ years. Thus, we
suggest that the increasing trend of the blue-MS stars' radial profile
may be associated with intensive binary dissolution in the cluster
cores.

The observed spatial distributions suggest that blue-MS stars are
partly soft binaries, which is at odds with the rotation-spread
scenario that interprets them as hard binaries. Our work places new
constraints on the origin of blue-MS stars. We note that the
classification of the three subpopulations, the radial bin size, and
the selected field stars may affect the accuracy of the observed
radial profiles. The external environment may affect the spatial
distributions of the stellar components. More information is needed to
reveal the origin of blue-MS stars.

\acknowledgments 
We are grateful for the detailed suggestions from the anonymous
referee that have helped improve this manuscript. We would like to
thank Dr. Jincheng Yu for making the $N$-body simulations
available. Y. Y. gratefully acknowledges financial support from the
China Scholarship Council (grant 201906010218). L. D. acknowledges
research support from the National Natural Science Foundation of China
through grants 11633005, 11473037, and U1631102. C. L. acknowledges
research support from the National Science Foundation of China through
grants 12037090.  C. L. and L. D. are grateful for support from the
National Key Research and Development Program of China through grant
2013CB834900 from the Chinese Ministry of Science and Technology. This
work is based on observations made with the NASA/ESA {\sl Hubble Space
  Telescope}, obtained from the data archive at the Space Telescope
Science Institute, which is operated by the Association of
Universities for Research in Astronomy, Inc. under NASA contract NAS
5-26555, under GO-14204, GO-14710, GO-13727, GO-14714, GO-14069,
GO-14204.

\software{DrizzlePac \citep{drizzlepac}, DOLPHOT \citep{dolphot2016},
  PARSEC \citep[1.2S;][]{Bressan2012}, Astropy \citep{astropy2013},
  Matplotlib \citep{Hunter2007}, PeTar \citep{2020MNRAS.497..536W},
  PySynphot \citep{pysynphot}}

\bibliography{ms}
\end{document}